\documentclass[apj]{emulateapj}
 \usepackage{CJK}
\usepackage{amsmath}
\usepackage{amssymb}
\usepackage{graphicx}
\usepackage{threeparttable}
\usepackage{bm}
\def\bfnabla{{\mbox{\boldmath $\nabla$}}}

\renewcommand\bv{{\mbox{\boldmath $v$}}}
\newcommand\bb{{\mbox{\boldmath $B$}}}
\newcommand\bP{{\mbox{\boldmath $P$}}}
\newcommand\bF{{\mbox{\boldmath $F$}}}
\newcommand\bfr{{\sf\boldmath f}}
\newcommand\bI{{{\sf\boldmath I}}}

\newcommand\Crat{{\mathbb{C}}}
\newcommand\Prat{{\mathbb{P}}}
\def\<{\,\langle\langle}
\def\>{\,\rangle\rangle}

\begin{document}

\begin{CJK*}{UTF8}{gbsn}

\title{Saturation of the MRI in Strongly Radiation Dominated Accretion Disks}

\shortauthors{Y.-F. Jiang et al.}

\author{Yan-Fei Jiang(姜燕飞)\altaffilmark{1}, James M. Stone\altaffilmark{1} \& Shane W. Davis\altaffilmark{2}}

\affil{$^1$Department of Astrophysical Sciences, Princeton
University, Princeton, NJ 08544, USA} 
\affil{$^2$Canadian Institute for Theoretical Astrophysics. Toronto, ON M5S3H4, Canada}

\begin{abstract}
The saturation level of the magneto-rotational instability (MRI) in a strongly radiation 
dominated accretion disk is studied using a 
new Godunov radiation MHD code in the unstratified shearing box 
approximation. 
Since vertical gravity is neglected in this work, 
our focus is on how the MRI saturates in the 
optically thick mid-plane of the disk. 
We confirm that turbulence generated 
by the MRI is very compressible in the radiation 
dominated regime, as found by previous calculations using the flux-limited diffusion approximation. 
We also find little difference in the saturation properties in calculations that use 
a larger horizontal domain (up to four times the vertical scale height in the radial direction). 
However, in strongly radiation pressure dominated disks (one in which the radiation 
energy density reaches $\sim 1\%$ of the rest mass energy density of the gas), we find 
Maxwell stress from the MRI 
turbulence is larger than the value produced when radiation pressure is replaced 
with the same amount of gas pressure. At the same time, the ratio between 
Maxwell stress and Reynolds stress is increased by almost a factor of $8$ compared 
with the gas pressure dominated case.  We suggest that this effect 
is caused by radiation drag, which acts like bulk viscosity and 
changes the effective magnetic Prandtl 
number of the fluid. Radiation viscosity significantly exceeds both the microscopic plasma
viscosity and resistivity, ensuring that radiation dominated systems occupy the high magnetic Prandtl 
number regime. Nevertheless, we find radiative shear viscosity is negligible compared to 
the Maxwell and Reynolds stress in the flow. 
This may have important implications for the structure of 
radiation dominated accretion disks. 
%We also find the tilte angle, or the 
%ratio between Maxwell stress and magnetic pressure is also smaller for the radiation 
%dominated case compared with the pure gas pressure case. 
% However, if we measure the ratio between 
%the Maxwell stress and gas pressure, it is roughly a constant value when the total 
%pressure is increased.
%radiation bulk viscosity. 

\end{abstract}

\keywords{(magnetohydrodynamics:) MHD $-$ methods: numerical $-$  radiative transfer}

\maketitle

\section{Introduction}
The inner regions of accretion disks around compact object are expected to be 
radiation pressure dominated \citep[][]{ShakuraSunyaev1973,Pringle1981}. Near the Eddington limit, 
the radiation field not only dominates the energy budget of the accretion 
disk, but also carries a significant fraction of the total momentum. 
The large accretion rate required to explain the observed X-ray luminosities of 
some compact objects cannot be explained by ordinary particle 
viscosity. Because the physical process to transfer angular momentum was 
unknown, \cite{ShakuraSunyaev1973} proposed the $\alpha$ disk model 
based on the assumption that the stress scales with the vertically integrated total 
pressure with a constant coefficient $\alpha$. 
Magneto-rotational instability (MRI) is now generally believed to be the 
physical mechanism to produce the required stress 
in ionized accretion disks \citep[e.g.,][]{BalbusHawley1991,BalbusHawley1998}. 
For the case with pure gas pressure, 
numerical simulations of MRI \citep[e.g.,][]{HGB1995,Stoneetal1996,MillerStone2000} show 
that it saturates as a turbulent state and the Maxwell stress 
from the turbulence can reach $\sim 1\%$ --- $10\%$ of the gas pressure. 
It is therefore important to understand how the MRI saturates in a radiation 
dominated flow. In particular, when the radiation energy density becomes 
a non-negligible fraction of the rest mass energy density of the gas, 
the properties of the MRI turbulence in 
the mid-plane of accretion disks may be affected by radiation, and  
radiative viscosity may contribute to the transport of angular momentum in this regime \citep[e.g.,][]{LoebLaor1992}.

%In radiation-dominated disks, it is well known that the $\alpha$ prescription 
%leads to the thermal \citep[][]{ShakuraSunyaev1976}
%and viscous instability \citep[e.g.,][]{lightmanEardley1974} if the accretion 
%stress is assumed to be proportional to the total pressure. 
% If instead, the accretion stress is assumed to be proportional 
%to only the gas pressure, the disk is stable 
%\citep[][]{SakimotoCoroniti1981,Coroniti1981,Burm1985}. 
%One key step to improve disk models is to understand how the 
%stress depends on the gas and radiation pressure. 

A first step in studying the MRI in radiation pressure 
dominated disks was made by \cite{Turneretal2003}, who 
studied MRI in the mid-plane of radiation dominated 
accretion disk using a flux-limited diffusion (FLD) module 
in ZEUS \citep[][]{TurnerStone2001} and neglecting vertical 
gravity. The vertical structure of a stratified radiation dominated accretion 
disk was subsequently studied by \cite{Turner2004}, \cite{Hiroseetal2009} and 
\cite{Blaesetal2011}.
%. More recently, 
%\cite{Hiroseetal2009} studied the same problem, using the same code, but 
%with significantly reduced energy error. Both found that when the 
%accretion stress is calculated self-consistently from MRI turbulence, radiation 
%dominated accretion disks are thermally stable. 
One limitation of FLD is that it is based on single moment
closure, and it drops radiation inertia in the momentum equation. 
When the accretion rate is near the Eddington limit, the photon 
momentum is a significant fraction of the fluid momentum, the FLD 
closure may not be appropriate to model the flow.

Recently, we have developed an improved algorithm for 
radiation MHD (see 
\citealt[][thereafter JSD12]{Jiangetal2012} and \citealt[][]{Davisetal2012}), 
based on a two-moment closure that solves the time dependent 
radiation momentum equations, and therefore retain radiation inertia. 
In this paper we use this algorithm to study the saturation of the MRI 
in the radiation dominated accretion disks.
As a first step, we focus on the mid-plane of an accretion 
disk neglecting vertical gravity. 
\cite{Turneretal2003}
found that for the parameters corresponding to the 
radiation dominated regime of the standard $\alpha$ 
disk model, the turbulence generated by MRI 
becomes highly compressible, consistent with expectation from the linear analysis
\citep{BlaesSocrates2001}. The 
Maxwell stress was found to be very similar to the value from 
simulations with radiation pressure replaced 
by the same amount of gas pressure. 
Here we attempt to reproduce these results without 
invoking the FLD approximation, as well as extend the study 
to higher ratios of radiation to gas pressure. The effects of the 
box size on the MRI turbulence will also be addressed \citep[e.g.,][]{Bodoetal2008}.
In the case of gas pressure only, the properties of turbulence 
generated by the MRI have been studied extensively. 
For example, the ratio between 
Maxwell and Reynolds stress is always found 
to be $\sim 4-5$ \cite[][]{HGB1995,Blackmanetal2008}, 
and Maxwell stress is also found to scale with the 
magnetic pressure very well \citep[e.g.,][]{Blackmanetal2008,Guanetal2009,Hawleyetal2011,Sorathiaetal2012}. 
Our goal is to investigate how these properties of 
MRI turbulence are affected by radiation, 
especially in a strongly radiation dominated flow. 

This paper is organized as follows. 
In \S\ref{sec:equations}, we describe the 
equations we solve and the shearing 
periodic boundary condition we use, 
especially for the radiation quantities. 
Then in \S\ref{sec:Turner}, we repeat 
and extend the fiducial simulations 
done by \cite{Turneretal2003}. 
The saturation states of MRI for 
different total pressure are studied 
in \S\ref{sec:diffT}. Discussions and 
conclusions are given in \S\ref{sec:discussion}.

%Then vertical structure of the radiation dominated accretion 
%disk is studied by \cite{Turner2004}.
  %

%Study how MRI will be changed by strong radiation field.

%In the standard $\alpha$ disk model \citep[e.g.,][]{ShakuraSunyaev1973}, 
%the stress, which is responsible for the accretion, is assumed to be 
%proportional to the total pressure. In other words, $\alpha$ is assumed to be 
%a constant. This parameterization is adopted because the fundamental 
%mechanism to provide the accretion stress is unknown at that time. As 
%MRI is believed to be responsible for the accretion stress, there is no 
%guarantee that the Maxwell stress and Reynolds stress generated by the 
%MRI turbulence will scale with the total pressure. 

\section{Equations}
\label{sec:equations}
We adopt the local shearing box approximation with shearing 
periodic boundary conditions  \citep[e.g.,][]{HGB1995} for this work. 
%This approximation has been widely used to study MRI, both with and 
%without radiation field \citep[e.g.,][]{Stoneetal1996,Turneretal2003,Bodoetal2008,Sanoetal2004,Davisetal2010}.  
In this approximation, we solve the equations of motion in a 
frame rotating with orbital frequency $\Omega$ at a fiducial radius from the central black hole. 
Curvature is neglected and the equations are written in
local Cartesian coordinates $(x,y,z)$ with unit vectors ($\bm{\hat{i}}$, $\bm{\hat{j}}$, $\bm{\hat{k}}$), 
which represent the radial, azimuthal and vertical direction respectively. 
Tidal and Coriolis forces are included in the local frame. 

The implementation of the local shearing 
box approximation in Athena without radiation are described in detail by \cite{StoneGardiner2010}. For radiation MHD, 
we need to add radiation source terms as described by JSD12. 
The radiation MHD equations in the mixed frame 
under the local shearing box approximation are 
\begin{eqnarray}
\frac{\partial\rho}{\partial t}+\bfnabla\cdot(\rho \bv)&=&0, \nonumber \\
\frac{\partial( \rho\bv)}{\partial t}+\bfnabla\cdot({\rho \bv\bv+{{\sf P}^{\ast}}}) &=&-{\bf \tilde{S}_r}(\bP)\nonumber \\
&+&2\rho\Omega^2qx\bm{ \hat i}-2\Omega \bm{ \hat k}\times \rho\bv,\  \nonumber \\
\frac{\partial{E}}{\partial t}+\bfnabla\cdot\left[(E+P^{\ast})\bv-\bb(\bb\cdot\bv)\right]&=&-c\tilde{S}_r(E)\nonumber\\
&+&\Omega^2\rho\bv\cdot(2qx\bm{ \hat i}),  \nonumber \\
\frac{\partial\bb}{\partial t}-\bfnabla\times(\bv\times\bb)&=&0, \nonumber \\
\frac{\partial E_r}{\partial t}+\bfnabla\cdot \bF_r&=&c\tilde{S}_r(E), \nonumber \\
\frac{1}{c^2}\frac{\partial \bF_r}{\partial t}+\bfnabla\cdot{\sf P}_r&=&{\bf \tilde{S}_r}(\bP).
\label{dimequation}
\end{eqnarray}
In the above equations, the shear parameter $q\equiv -d\ln\Omega/d\ln r$ and 
$q=3/2$ for Keplerian rotation while 
$\rho$ is density, ${\sf P}^{\ast}\equiv(P+B^2/2)\bI$ (with $\bI$
the unit tensor), 
and the magnetic permeability $\mu=1$.  The total gas energy density is
\begin{eqnarray}
E=E_g+\frac{1}{2}\rho v^2+\frac{B^2}{2},
\end{eqnarray}
where $E_g$ is the internal gas energy density.   We adopt the equation of state
for an ideal gas with adiabatic index $\gamma=5/3$. Then
$E_g=P/(\gamma-1)$ and $T=P/R_{\text{ideal}}\rho$, where
$R_{\text{ideal}}$ is the ideal gas constant. 
Radiation pressure, energy density and flux are ${\sf P}_r, \ E_r$ and $ \bF_r$ 
respectively while $c$ is the speed of light. The radiation momentum and energy source terms 
are ${\bf\tilde{S}_r}(\bP)$ and $\tilde{S}_r(E)$.

Following \cite{Jiangetal2012}, we use a dimensionless set of equations and variables
in the remainder of this work.  We convert the above set of equations to the dimensionless 
form by choosing fiducial units for velocity, density, temperature and 
pressure as $a_0,\ \rho_0,  \ T_0$ and $P_{g,0}$ respectively. A dimensionless 
parameter $\Prat$ can be defined as $\Prat\equiv a_rT_0^4/P_{g,0}$, where 
$a_r$ is the radiation constant. 
Then units for radiation energy density $E_r$ and flux $\bF_r$ are $a_rT_0^4$ 
and $ca_rT_0^4$. In other words, $a_r=1$ in our units. The dimensionless speed of light is $\Crat\equiv c/a_0$. 
The original dimensional equations can then be written to the following dimensionless form 
\begin{eqnarray}
\frac{\partial\rho}{\partial t}+\bfnabla\cdot(\rho \bv)&=&0, \nonumber \\
\frac{\partial( \rho\bv)}{\partial t}+\bfnabla\cdot({\rho \bv\bv-\bb\bb+{{\sf P}^{\ast}}}) &=&-\mathbb{P}{\bf S_r}(\bP)\nonumber\\
&+&2\rho\Omega^2qx\bm{ \hat i}-2\Omega \bm{ \hat k}\times \rho\bv,\  \nonumber \\
\frac{\partial{E}}{\partial t}+\bfnabla\cdot\left[(E+P^{\ast})\bv-\bb(\bb\cdot\bv)\right]&=&-\mathbb{PC}S_r(E)\nonumber\\
&+&\Omega^2\rho\bv\cdot(2qx\bm{ \hat i}),  \nonumber \\
\frac{\partial\bb}{\partial t}-\bfnabla\times(\bv\times\bb)&=&0, \nonumber \\
\frac{\partial E_r}{\partial t}+\mathbb{C}\bfnabla\cdot \bF_r&=&\mathbb{C}S_r(E), \nonumber \\
\frac{\partial \bF_r}{\partial t}+\mathbb{C}\bfnabla\cdot{\sf P}_r&=&\mathbb{C}{\bf S_r}(\bP),
\label{equations}
\end{eqnarray}
where the source terms are,
\begin{eqnarray}
{\bf S_r}(\bP)&=&-\left(\sigma_{aF}+\sigma_{sF}\right)\left(\bF_r-\frac{\bv E_r+\bv\cdot{\sf P} _r}{\mathbb{C}}\right)\nonumber \\
&+&\frac{\bv}{\mathbb{C}}(\sigma_{aP}T^4-\sigma_{aE}E_r),\nonumber\\
S_r(E)&=&(\sigma_{aP}T^4-\sigma_{aE}E_r) \nonumber \\
&+&(\sigma_{aF}-\sigma_{sF})\frac{\bv}{\mathbb{C}}\cdot\left(\bF _r-\frac{\bv E_r+\bv\cdot{\sf  P} _r}{\mathbb{C}}\right).
\label{sources}
\end{eqnarray}
Frequency mean absorption and scattering opacities (attenuation coefficients) are 
$\sigma_{aF}$, $\sigma_{sF}$ respectively while $\sigma_{aP}$ and $\sigma_{aE}$ are Planck mean 
and energy mean absorption opacity (attenuation coefficients). To change the dimensionless equations 
to the dimensional form, we only need to replace $\Crat$ with $c$, set $\Prat=1$ 
and replace $\bF_r$ with $\bF_r/c$. Note that we solve the time-dependent radiation momentum 
equations, and therefore include radiation inertia effect. 

The radiation pressure ${\sf P}_r$ is related to the radiation energy density $E_r$
with the variable Eddington tensor $\bfr$:
\begin{eqnarray}
{\sf P}_r=\bfr E_r.
\end{eqnarray}
In principle, $\bfr$ should be calculated with our short characteristic module as 
described by \cite{Davisetal2012}. We have tried this approach 
and find that the difference between the Variable Eddington tensor (VET) 
and $1/3\bI$ is smaller than $10^{-4}$. 
Thus for unstratified disk simulations with 
uniform density in the optically thick regime, the Eddington approximation is 
adequate, and we will adopt it in this paper. 
%Eddington approximation is applied in the Eulerian frame Its effect 
%on co-moving frame will be discussed later

We use the recently developed Godunov radiation MHD method (JSD12) to solve those 
equations. This algorithm has been implemented and tested in Athena \citep[][]{Stoneetal2008} as 
described in JSD12 with several improvements as explained in Appendix \ref{codechange}. 
The momentum and energy source terms resulting from the shearing box
approximation are added in the same way as  in \cite{StoneGardiner2010}, which are 
separated from the radiation source terms.  

\subsection{Boundary Condition}
We adopt the usual shearing periodic boundary condition in $x$, and 
periodic boundary condition in both $y$ and $z$ directions \citep[e.g.,][]{HGB1995}. 
Boundary conditions for the gas quantities are the same as in non-radiative MHD 
simulations \citep[e.g.,][]{StoneGardiner2010}. The radiation quantities 
$E_r$ and $\bF_r$ are new variables, and boundary conditions for them 
deserve detailed discussion. 

If the radial size of the simulation box is $L_x$, then the boundary condition for $E_r$ 
is 
\begin{eqnarray}
E_r(x,y,z)\mapsto E_r(x\pm L_x,y\mp q\Omega L_x t,z).
\end{eqnarray}
This boundary condition is the same as the boundary condition for density. $\bF_r$ 
is the Eulerian radiation flux, which includes the co-moving and 
advected radiation flux. We first remap the co-moving radiation flux and then
add the advection flux at the new location. Thus the boundary condition 
for $\bF_r$ is 
\begin{eqnarray}
F_{r,x}&\mapsto& F_{r,x}(x\pm L_x,y\mp q\Omega L_x t,z), \nonumber\\
F_{r,y}&\mapsto& F_{r,y}(x\pm L_x,y\mp q\Omega L_x t,z) \nonumber \\
&\mp& (1+f_{yy})\frac{q\Omega L_x}{\Crat}E_r(x\pm L_x,y\mp q\Omega L_x t,z), \nonumber \\
F_{r,z}&\mapsto& F_{r,z}(x\pm L_x,y\mp q\Omega L_x t,z),
\end{eqnarray}
where $f_{yy}$ is the $yy$ component of the Eddington tensor at the new location.

\subsection{Energy Conservation}
Unlike the ZEUS code, Athena conserves energy to roundoff error when energies
exchange between kinetic, internal and magnetic forms. 
However, as explained in JSD12, when photons and gas exchange energy, 
the radiation module in Athena does not conserve total 
energy to round off error because of our splitting scheme and the  
implicit matrix solver. With the special treatment of the energy 
source term described in JSD12, the energy error is kept small. 
This is particularly important for MRI simulations because 
the gas is continuously heated up during the simulations. If the energy error 
is not treated carefully, it can accumulate and 
eventually affect the dynamics. 

For unstratified simulations with shearing periodic boundary conditions, 
the change of the total energy inside the simulation 
box is entirely due to the work done on the walls \citep[e.g.,][]{HGB1995,GardinerStone2005}. 
With radiation, the volume averaged total energy is $E_t\equiv\langle E+\rho\Phi+\Prat E_r\rangle$, 
where $\Phi$ is the tidal potential $\Phi=-q\Omega^2x^2$. The change of 
total energy is determined by
\begin{eqnarray}
\frac{\partial E_t}{\partial t}=\frac{q\Omega}{L_yL_z}\int_X
\left(\rho v_x\delta v_y-B_xB_y\right)dydz\equiv W(t),
\label{wallwork}
\end{eqnarray}
where $\delta v_y\equiv v_y+q\Omega x$, and the integral only needs to be done on one 
side of the radial boundary. The total work done on the simulation box within a certain time 
$t_0$ is then $W_{t}(t_0)=\int_0^{t_0} W(t)dt$. To check the energy error, we calculate the 
energy change at $t_0$ as $\delta E=E_t(t_0)-E_t(0)$, so that the energy error is $\delta E-W_{t}(t_0)$.
We calculate this energy error for all simulations and find that with respect 
to the total work $W_{t}(t_0)$ within the first $100$ orbits, the energy errors are always smaller than $1\%$.

\section{Results}
In this section, we first repeat the fiducial simulations 
done by \cite{Turneretal2003} and then extend them to a box with a larger radial size. 
%This can be used as tests of our code 
%in the diffusion regime. 
In addition, we will study the saturation state 
of MRI turbulence with different radiation pressure. 

\subsection{Simulations with Net Vertical Flux}
\label{sec:Turner}
\cite{Turneretal2003} used the FLD module in ZEUS \citep[][]{TurnerStone2001} 
to study MRI turbulence in the optically thick regime of a 
unstratified disk model. The radial size of the simulation box 
was fixed to be one scale height. 
Here we repeat their 
simulations with net vertical flux to compare the results from the two different codes. 
Because the whole simulation box is very optically thick, we expect to find similar results as 
those computed with FLD. 
We also extend the radial size of the simulation box to four scale height to study 
possible new effects with radiation in this larger radial domain \citep[e.g.,][]{Bodoetal2008}. 

\subsubsection{Initial Condition}

We take the parameters for location I in table 1 of \cite{Turneretal2003}. 
This corresponds to a radius $67.8r_G$ from a $10^8M_{\odot}$ supermassive black 
hole, where the gravitational radius $r_G\equiv GM/c^2$ and $M$ is the black hole mass. 
The whole simulation box has uniform density $\rho_0=2.89\times 10^{-9}$ g cm$^{-3}$ and
temperature $T_0=2.71\times 10^5$ K. The density and temperature correspond to the 
mid-plane in an $\alpha$ disk model with $\alpha=0.01$ 
and accretion rate $10\%$ of the Eddington rate. The  
mean molecular weight is assumed to be $0.6$. Gas 
and radiation are in thermal equilibrium initially. We include both electron scattering opacity 
$\kappa_{\rm T}=0.4$ cm$^2$ g$^{-1}$ and free-free absorption opacity $\kappa_{\rm ff}=10^{52}\rho^{9/2}\left[P_g/(\gamma-1)\right]^{-7/2}$ 
\text{cm$^2$ g$^{-1}$}. Taking the unit of density to be $\rho_0$, temperature to be $T_0$, 
pressure to be $P_{g,0}=\rho_0 R_{\text{ideal}}T_0$ (with $R_{\text{ideal}}$ to be the 
ideal gas constant), 
time to be $1/\Omega$, and length to be $L_0=(P_{g,0}/\rho_0)^{0.5}/\Omega$, the dimensionless parameters in our 
equations are $\Prat=376.3$, $\Crat=4895.6$, $\sigma_{sF}=1953.9$, and $\sigma_{aF}=\sigma_{aP}=\sigma_{aE}=0.105$. 
Thus the ratio between the background radiation and gas pressure is $\sim 125$. 
The ratio between radiation energy density and rest mass energy of the fluid is 
$a_rT_0^4/\left(\rho_0c^2\right)=3.14\times10^{-5}$. Table \ref{ZFtable} lists other parameters for 
the five simulations in this section, with different box sizes and resolutions.

The initial magnetic field is $B_x=B_y=0,B_z=0.0353$. The most unstable MRI wavelength 
is $\lambda_z=2\pi v_{A,z}/\Omega=2\pi\sqrt{B^2/\rho}/\Omega=0.22L_0$. 
Notice that with the unit of length we choose, the size of the simulation box are smaller than the fiducial 
runs done by \cite{Turneretal2003}. The net magnetic vertical flux we use is also smaller so that 
the ratio between the most unstable MRI wavelength and the vertical 
size of the simulation box is the same as in \cite{Turneretal2003}. 
We have found that choosing a smaller field strength produces less 
vigorous channel solutions \citep[e.g.,][]{GoodmanXu1994} and leads to 
a more rapid transition to turbulence. 
Initial random perturbations 
are applied to both gas pressure and velocity field
\begin{eqnarray}
P_g=P_{g,0}(1+\delta_1), v_x=\delta_2, v_y=\delta_3-q\Omega x, v_z=\delta_4,
\label{perturbation}
\end{eqnarray}
where $\delta_1$ is a random number uniformly distributed between $-0.1$ and $0.1$ while 
$\delta_2,\delta_3,\delta_4$ are random numbers uniformly distributed between $-0.02$ and $0.02$. 
Athena's orbital advection scheme, as described in \cite{StoneGardiner2010}, is used for all the simulations 
to speed up the calculation. For the radiation field, the advective radiation flux due to the mean orbital 
motion is separated from other terms and added as described in 
Appendix \ref{AdvEr}.

\begin{table*}[htdp]
\centering
\begin{threeparttable}
\caption{Parameters for simulations with net vertical flux}
\begin{tabular}{ccccccccc}
\hline
Label & $L_x/L_0$ & Grids/$L_0$ & $\tau_s $ & $\tau_a$ & $P_{g}$ & $P_{r}$ & $B_{z,0}^2/2$  \\
\hline
ZFRS & 1 & 32  	&  1954	&  0.11				&  1		& 125 &	$6.25\times10^{-4} $	 \\
ZFRL & 4 & 32  	&   1954	&  0.11				&  1		& 125 &	$6.25\times10^{-4} $	 \\
ZFNRS & 1 & 32  	&   ---	&  ---				&  126	&  ---	& 	$6.25\times10^{-4} $		 \\
ZFNRL & 4 & 32  	&  ---		&  ---				& 126	&  --- &  $6.25\times10^{-4} $ 		 \\
ZFRS64 & 1 & 64  	&  1954	&  0.11				&  1		& 125 & $6.25\times10^{-4} $ 	 \\
%ZFRL64 & 4 & 64  	&  1954	&  0.11				& 1		&125 &  $6.25\times10^{-4} $		\\
%ZFThin & 4 & 64 	&  26.36	& $9.8\times 10^{-5}$	& 1		& 125 &  $6.25\times10^{-4} $		\\
\hline
%\multicolumn{8}{l}{{Notes: $\tau_s$ is the electron scattering optical depth per $H$. }} \\
%\multicolumn{8}{l}{{$\tau_a$ is the 
%Planck mean free-free absorption optical depth per $H$. }}
\end{tabular}
\begin{tablenotes}
\item [1]	$\tau_s$ is the electron scattering optical depth per $L_0$.
\item [2]	$\tau_a$ is the Planck mean free-free absorption optical depth per $L_0$.
\end{tablenotes}
\label{ZFtable}
\end{threeparttable}
\end{table*}%

\begin{table*}[htdp]
\caption{Temporal and Spatial Averaged Properties of Simulations with Net Vertical flux}
\centering
\begin{tabular}{cccccccccc}
\hline
Label & $\<-B_xB_y\>$ & $\<\rho v_x\delta v_y\>$ & $\<B^2/2\>$ & $\<B_x^2/2\>$ & $\alpha_{\text{stress}}$ \\
\hline
ZFRS	&	0.154	&	0.0282	&	0.320	&	0.0398	&  	5.46	\\
ZFRL	&	0.162	&	0.0291	&	0.338	&	0.0530	&	5.57	\\
ZFNRS	&	0.235	&	0.0347	&	0.518	&	0.0560      &  	6.77	\\
ZFNRL	&	0.123	&	0.0273	&	0.240	&	0.0390	&	4.51	\\
ZFRS64	&	0.321	&	0.0486	&	0.761	&	0.137	&	6.71	\\
%ZFRL64	&	0.1639	&	0.0288	&	0.3553	&	0.0562	&	\\
%ZFThin	&	0.1035	&	0.0242	&	0.2059	&	0.0370	&	\\
\hline
\hline
Label &  $\<B_y^2/2\>$ & $\<B_z^2/2\>$ & $\<T/T_0\>$  &  $\alpha_{\text{mag}} $	& $\alpha $\\
\hline
ZFRS	&	0.268	&	0.0113	&		1.04 		&	0.481 		&	0.00145	\\
ZFRL	&	0.263	&	0.0220	&		1.05		&	0.478		&	0.00151	\\
ZFNRS	&	0.448	&	0.0150	&		1.61		&	0.453		&	0.00214	\\
ZFNRL	&	0.188	&	0.0130	&		1.41		&	0.513		&	0.00119	\\
ZFRS64	&	0.556	&	0.0683	&		1.09		&	0.428		&	0.00297	\\
%ZFRL64	&	0.2761	&	0.0230	&		1.0531	&	0.4613		&	0.08106	\\
%ZFThin	&	0.1538	&	0.0151	&		1.0371	&	0.5027		&	0.1175	\\
\hline
\hline
\end{tabular}
\label{ZFtable2}
\end{table*}%

\subsubsection{Time Evolution}
The fiducial simulation in Table \ref{ZFtable}, ZFRS, has 
the standard geometry $L_x=L_0, L_y=4L_0, L_z=L_0$. To see the 
effect of the size of the simulation domain, we increase the radial 
size such that $L_x=4L_0$ and label this simulation ZFRL. For comparison, 
the simulations ZFRNS and ZFNRL have the same parameters as ZFRS 
and ZFRL respectively, except that 
radiation pressure is replaced with the same amount of gas pressure. 
The differences between the two sets of simulations will be due to the radiation field. 
Simulation ZFRS64 has double the resolution of ZFRS. 
 
\begin{figure}[htp]
\centering
\includegraphics[width=0.96\hsize]{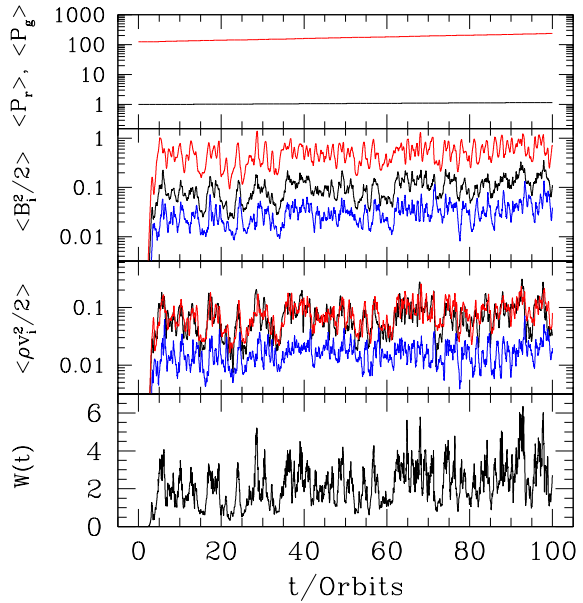}
\caption{Time history of the spatially averaged quantities from simulation ZFRS with 
net vertical flux. The radiation (the red line) and gas (the black line) pressure are 
shown in the top panel. Components of the magnetic and kinetic energy densities 
are shown in the middle two panels. The black, red and blue lines represent $x$, $y$ 
and $z$ components respectively. Background shearing is not included in the 
calculation of kinetic energy density. The bottom panel shows the total work done on the 
wall per unit time, calculated according to equation~\ref{wallwork}. }
\label{ZFRSHistory}
\end{figure}

\begin{figure*}[htp]
\centering
\includegraphics[width=0.49\hsize]{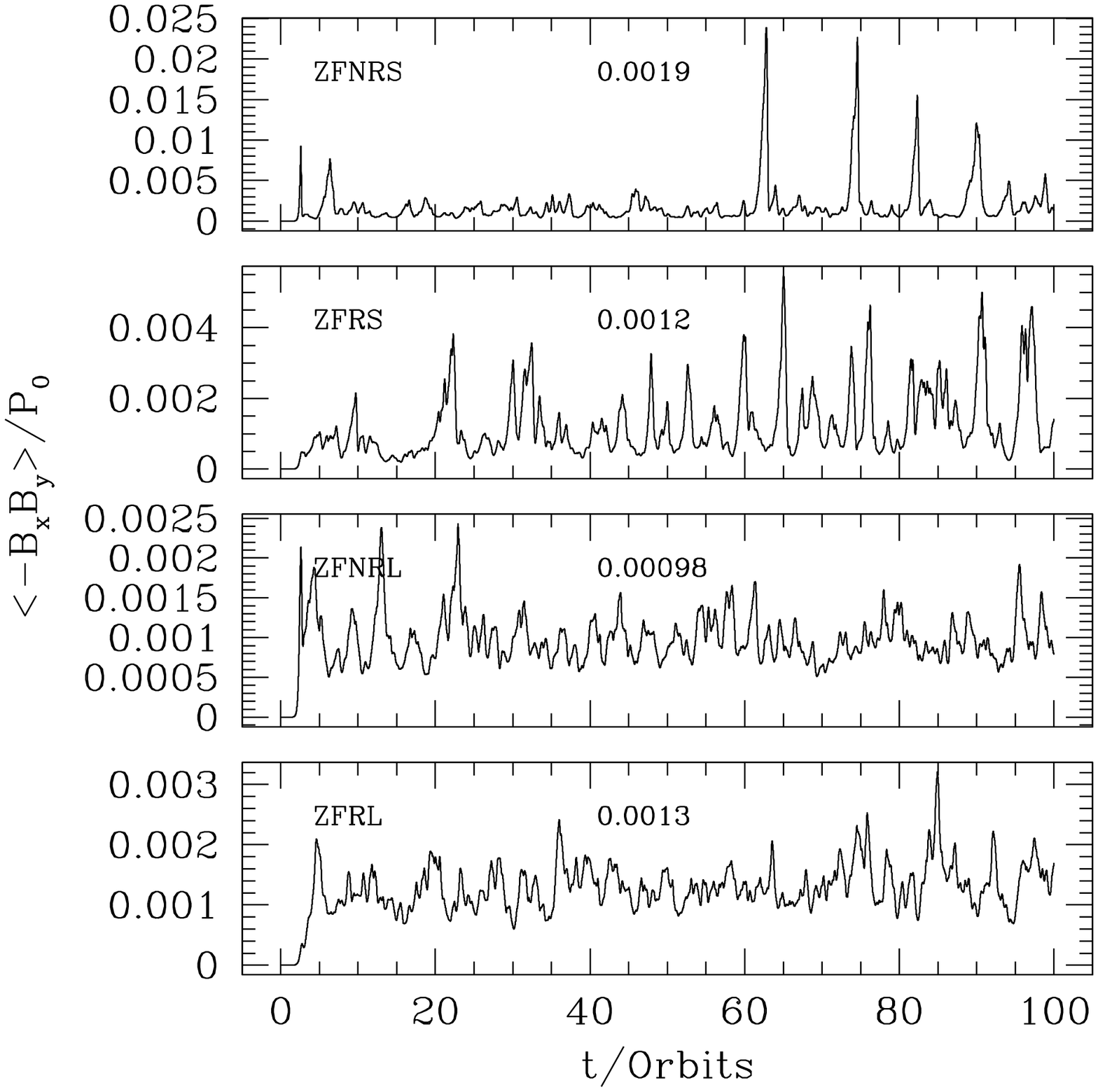}
\includegraphics[width=0.49\hsize]{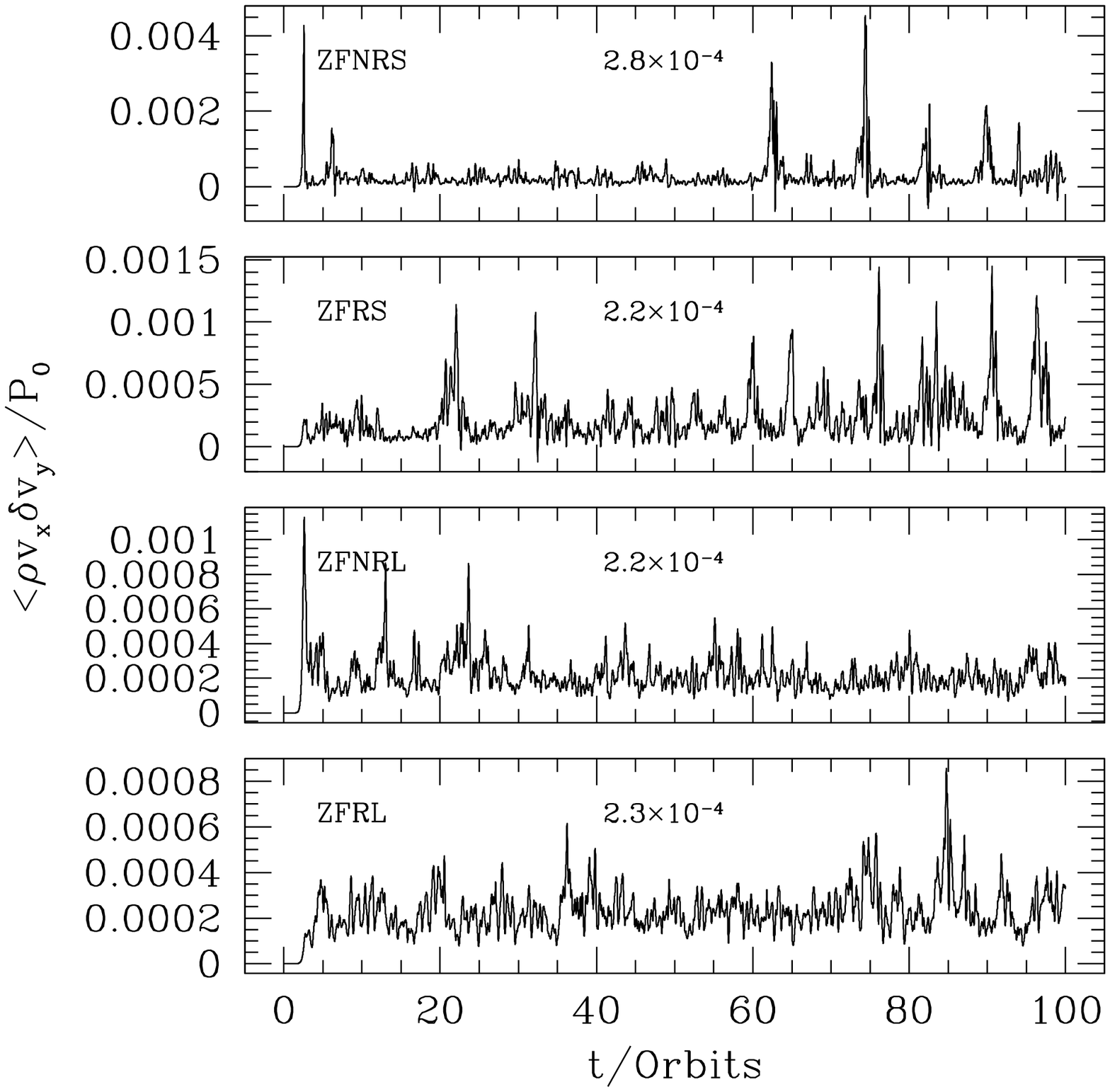} 
\caption{History of the spatially averaged Maxwell stress (left panels) and Renolds stress (right panels) 
for the four simulations with net vertical flux. Parameters of the four simulations are given in 
Table \ref{ZFtable}. The stress is normalized with respect to the initial total pressure. The 
number in each panel is the time averaged value between 10 and 100 orbits. }
\label{Turnerstress}
\end{figure*}

In Figure \ref{ZFRSHistory}, we show the time evolution 
of radiation and gas pressure, three components of magnetic 
and kinetic energy densities as well as the total work 
done on the wall of the simulation box per unit time for 
the fiducial run ZFRS. Because photons cannot leave the system in the 
unstratified shearing box models, and moreover the box is continuously 
heated up due to the work done on the walls, there is no equilibrium solution 
and radiation and gas pressure continue to increase. 
Since $E_r\propto T^4$ while gas internal energy density $E_g\propto T$, radiation 
energy density increases faster and most of the work is converted to 
the radiation energy density. Although total energy cannot reach an equilibrium 
state, the magnetic and kinetic energy densities saturate quickly after 
the initial transient. They fluctuate around some mean values, 
which do not show any systematic change with pressure within the $100$ orbits. 
This is also true for the Maxwell and Reynolds stress, 
the histories of which for four simulations are shown in 
Figure \ref{Turnerstress}. 
%The behaviors of these quantities 
%are consistent with \cite{Turneretal2003}. 
Time histories of these quantities 
are very similar for simulations with or without radiation and it is 
independent of the box size.

\subsubsection{Statistical Properties}

In Table \ref{ZFtable2}, we calculate different spatial and temporal averaged 
quantities for the five simulations. The temporal averages are taken between 
$10$ and $100$ orbits to avoid the initial transient before the MRI saturates. 
In that table, $\alpha_{\text{stress}}, \alpha_{\text{mag}} $ and 
$\alpha$ are defined as
\begin{eqnarray}
\alpha_{\text{stress}}&\equiv& \frac{\<-B_xB_y\>}{\<\rho v_x\delta v_y\>}, 
\alpha_{\text{mag}}\equiv \frac{\<-B_xB_y\>}{\<B^2/2\>}, \nonumber \\
\alpha&\equiv& \frac{\<-B_xB_y+\rho v_x\delta v_y\>}{P_0}, 
\end{eqnarray}
where $P_0$ is the initial total pressure and $\<\cdot\>$ represents spatial and 
temporal averaging. 

As shown in Table \ref{ZFtable2}, the averaged Maxwell stress, Reynolds stress and 
magnetic pressure are very similar for 
simulations ZFRS and ZFRL, which suggests that changing the box size 
along radial direction does not affect the statistical properties of the
turbulence. However, as shown in Figure \ref{Turnerstress}, when 
$L_x$ is increased by a factor of $4$, there are  fewer large spikes compared with the 
fiducial run. This is also true for the pure gas pressure runs
(simulations ZFNRS and ZFNRL). This is consistent with \cite{Bodoetal2008}
for MRI without radiation. The spikes in the fiducial run are due to the recurrent channel solutions. 
When the radial size is increased, the channel solution 
can be more easily destroyed either by the parasitic modes \citep[e.g.,][]{GoodmanXu1994,PessahGoodman2009}, 
or by the interactions between different MRI modes \citep[e.g.,][]{Latteretal2009}. 
%there are more parasitic modes 
%present inside the box, which destroy the channel solution \citep[e.g.,][]{GoodmanXu1994}. 

When radiation pressure is replaced with the same amount of gas pressure,  
Maxwell stress for simulation ZFNRS is about $50\%$ larger than for simulation ZFRS. 
The difference is primarily due to the large spikes
in the non-radiative run, which are weaker by almost a factor of $5$ for the simulation 
with radiation. If we exclude 
those spikes, ZFNRS has a very similar saturation level as ZFRS. 
The change in temperature is $50\%$ larger during the first $100$ orbits 
for ZFNRS than ZFRS, although the total work done on the fluid is very 
similar for the two simulations. %This is because radiation energy density $E_r\propto T^4$ 
%while gas internal energy $P_g/(\gamma-1)\propto T$. Thus 
This is because the heat capacity for the photons is much larger than an ideal gas. 
When $L_x$
is increased by a factor of $4$, ZFNRL and ZFRL have roughly the same 
Maxwell stress and Reynolds stress. This is consistent with Figure 2 of \cite{Turneretal2003}, 
which shows that the stress is very similar when the net vertical flux and total 
pressure are the same for their simulations NV4 and RV4. The stresses from 
our simulations are smaller than what \cite{Turneretal2003} found because we use 
a smaller net vertical flux. This is roughly consistent with the scaling between the 
saturated Maxwell stress and the initial net vertical magnetic flux as found by 
\cite{HGB1995}. Simulations RV4h, RV4 and RV4l in \cite{Turneretal2003} also 
have very similar stress, although the three simulations have quite different 
opacities and thus different coupling between the photons and gas. 

The ratio between 
Maxwell and Reynolds stress, $\alpha_{\text{stress}}$, varies from $4.5$ to $6.8$ for the first  
four simulations in Table \ref{ZFtable2}.  
This is also consistent with previous MRI simulations with \citep[e.g.,][]{Turneretal2003} or 
without \citep[e.g.,][]{Sanoetal2004,Davisetal2010} radiation. 
% may need to check the exact numbers
When resolution is doubled for simulation ZFRS64, the Maxwell stress 
is doubled while Reynolds stress is increased by $70\%$ compared with 
simulation ZFRS. But the ratio between 
Maxwell stress and magnetic pressure $\alpha_{\text{mag}}$ varies 
in the small range $0.43 \sim 0.51$ for all the five simulations. 
This is consistent with previous work, which found 
 $\alpha_{\text{mag}}$, or the tilt angle defined based on 
it \citep[e.g.,][]{Guanetal2009,Hawleyetal2011,Sorathiaetal2012}, to be 
roughly constant for converged simulations with different box sizes and
initial conditions \citep[e.g.,][]{Blackmanetal2008,Hawleyetal2011}. 
Most previous MRI simulations without radiation find $\alpha_{\text{mag}} \sim 0.5$ \citep[e.g.,][]{Guanetal2009}, 
although a range of $0.3 - 0.4$ is also reported by \cite{Hawleyetal2011} for the stratified disk simulations.

\subsubsection{Density fluctuation}
\begin{figure}[htp]
\centering
\includegraphics[width=0.9\hsize]{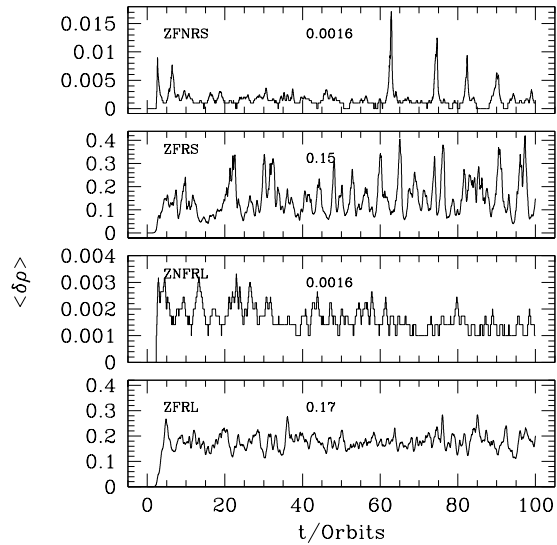}
\caption{Standard deviation of the density for the four simulations with net vertical flux listed in 
Table \ref{ZFtable}. The number in each panel is the time averaged value. The first 
and the third panels are for simulations without radiation for comparison, which have much 
smaller density fluctuation compared with the other two panels. }
\label{Turnersigmarho}
\end{figure}

When there is no radiation field, turbulence generated by the MRI is almost 
incompressible. However, 
\cite{Turneretal2003} finds that if the photon diffusion length 
per orbit is larger than the most unstable MRI wavelength, 
the turbulence generated by MRI with radiation field 
becomes very compressible. This phenomenon can be 
characterized by the standard deviation of density 
$\delta\rho\equiv\sqrt{\langle (\rho-\rho_0)^2 \rangle}$. For simulations 
ZFRS, ZFRL and ZFRS64, photon diffusion 
length per orbit $\left[2\pi\Crat/\left(3(\sigma_{aF}+\sigma_{sF})\Omega\right)\right]^{1/2} = 2.3$ 
while the most unstable MRI wavelength is only $0.25$. Thus we are in the regime in which
large density fluctuations are expected. Indeed, as shown in Figure \ref{Turnersigmarho}, 
for simulations without radiation, $\delta\rho$ is only $0.16\%\rho_0$ for both ZFNRS and ZFNRL. 
While for the simulations with radiation, $\delta\rho$ reaches $15\%\rho_0$ for ZFRS and $17\%\rho_0$ 
for ZFRL. 

In summary, we find that in the very optically thick regime, our results are 
similar to those computed using ZEUS with FLD for the parameters explored by 
\cite{Turneretal2003}. Changing the box size does not affect the statistical 
mean values of the turbulence, which is true for simulations with or without 
radiation. Although the total pressure can increase by a factor of 2 within $100$ 
orbits, saturation levels of Maxwell and Reynolds stress do not show a systematic 
trend of change with pressure in the unstratified disk simulations.

\subsection{Simulations with Net Toroidal Flux}
\label{sec:diffT}
%
%Because the thickness of the disk is fixed in the unstratified disk 
%models, effects of the disk scale height on the stress cannot be studied in the unstratified disk simulations. Nevertheless, 
%these simulations can still allow us to study the possible effects of 
%radiation pressure on the saturation level of MRI for a fixed disk scale height, as 
%studied by \cite{Sanoetal2004} for the pure gas pressure case. But we caution that 
%this may not be the dominant effect to determine the stress in real accretion 
%disks, which will be studied with stratified disk models in the future. 
In this section, we study the possible effects of radiation on the saturation 
states of the MRI for the cases with net toroidal flux. This configuration of 
magnetic field is commonly used for most stratified accretion disk 
simulations \citep[e.g.,][]{Hiroseetal2009}.

\subsubsection{Initial Condition}

Here we carry out a set of simulations with different temperature as listed in Table \ref{YFtable}. 
For these simulations, we choose our fiducial parameters to be the mid-plane 
values of the radiation dominated accretion disk studied by \cite{Hiroseetal2009}. 
The disk is located at $30r_G$ from a $6.62M_{\odot}$ black hole. The 
fiducial density and temperature are $\rho_0=5.66\times 10^{-2}$ g cm$^{-3}$ and
$T_0=2.63\times 10^7$ K. The orbital frequency $\Omega=190$ s$^{-1}$ and 
the time unit is $1/\Omega$. Mean molecular weight is assumed 
to be $0.6$ and the length unit is $L_0=\left(P_{g,0}/\rho_0\right)^{1/2}/\Omega$. 
The dimensionless parameters $\Prat=17.6$ and $\Crat=497$. We adopt the same 
opacities as used in \cite{Hiroseetal2009}. Electron scattering 
opacity is $0.33$ cm$^2$ g$^{-1}$, Planck and frequency mean free-free 
opacity are $3.7\times 10^{53}\rho^{9/2}\left[P_g/(\gamma-1)\right]^{-7/2}$ 
cm$^2$ g$^{-1}$ and  $10^{52}\rho^{9/2}\left[P_g/(\gamma-1)\right]^{-7/2}$ 
cm$^2$ g$^{-1}$ respectively. In our units, the dimensionless electron scattering 
and Planck mean absorption opacities (attenuation coefficients) with density $\rho_0$ 
and temperature $T_0$ are $\sigma_{sF}=5.94\times 10^3$ and
$\sigma_{aP}=31.05$. The box size is fixed to be $L_x=L_0$, $L_y=4L_0$ and $L_z=4L_0$. 

All the simulations listed in 
Table \ref{YFtable} start with a uniform density $0.5\rho_0$ but we change the 
initial temperature to setup different radiation pressures. Initial perturbations are 
applied according to equation \ref{perturbation}. For the fiducial parameters, 
the ratio between radiation energy density and rest mass energy of the fluid is 
$a_rT_0^4/\left(\rho_0c^2\right)=7.13\times10^{-5}$. 
For the case when temperature is $3T_0$, the ratio is increased to $5.78\times10^{-3}$.

Simulations YFR1, YFR2 and YFR3 start from thermal equilibrium states with 
temperature $T_0$, $2T_0$ and $3T_0$. 
The ratio between radiation pressure and gas pressure 
is $11.7$, $93.9$ and $317.0$ for the three simulations respectively. 
Two oppositely twisted magnetic flux tubes with the same azimuthal 
flux are initially placed at $-0.35 L_0<z<0.35 L_0$. $B_x$ and $B_z$ are initialized with the following 
vector potential $A_B$:
\begin{eqnarray}
A_B(x,y,z)=\left\{ \begin{array}
{r@{\quad:\quad}l}
0 & r > 1 \\  -\text{sign}(z)\frac{B_0}{4\pi} \left[ 
1+\cos\left(\pi r\right) \right] & r\leq 1
\end{array}
\right. ,
\end{eqnarray}
where 
\begin{eqnarray}
r\equiv \left\{ \begin{array}
{r@{\quad:\quad}l}
\left[
\left(x/L_x\right)^2+
(z-0.35L_0)^2/(0.35L_0)^2
\right]^{1/2} & z > 0 \\
\left[
\left(x/L_x\right)^2+
(z+0.35L_0)^2/(0.35L_0)^2
\right]^{1/2} & z < 0
\end{array}
\right. .
\end{eqnarray}
Then $B_y$ is initialized as 
$B_y=\left(B_0^2/2-B_x^2-B_z^2\right)^{1/2}$ 
such that magnetic pressure is uniform in the simulation box. 
We adopt $B_0^2/2=0.04$ for all the simulations in this 
section so that the net toroidal magnetic flux is always 
the same when we change the pressure. 
Simulation YFNR has the same setup as 
YFR1 except that radiation pressure is replaced with the 
same amount of gas pressure. 
%Simulations VFR1, VFR2 and VFR3 have similar parameters 
%as YFR1, YFR2 and YFR3, except that they are initialized with 
%pure vertical flux and simulation box elongates along $x$ 
%direction. 
Simulations YFR1Res, YFR2Res and YFR3Res 
are doubled resolution version of YFR1, YFR2 and YFR3
respectively.

\begin{figure}[htp]
\centering
\includegraphics[width=0.96\hsize]{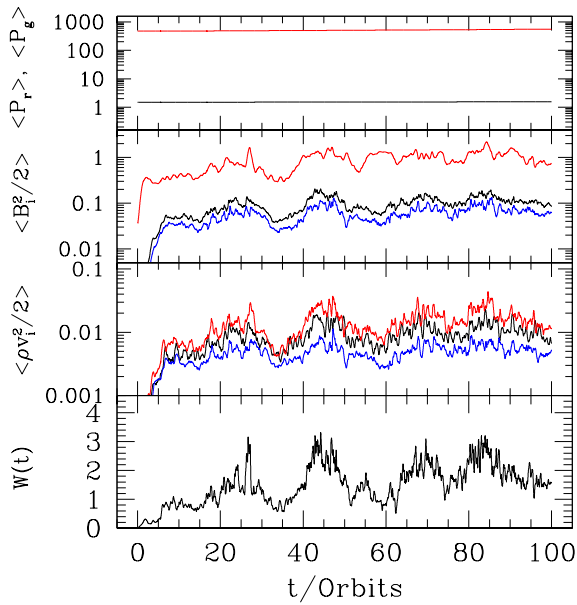}
\caption{The same as Figure~\ref{ZFRSHistory} but for simulation YFR3Res with net toroidal flux.}
\label{YFT3ResHistory}
\end{figure}

\begin{figure}[htp]
\centering
\includegraphics[width=1.2\hsize]{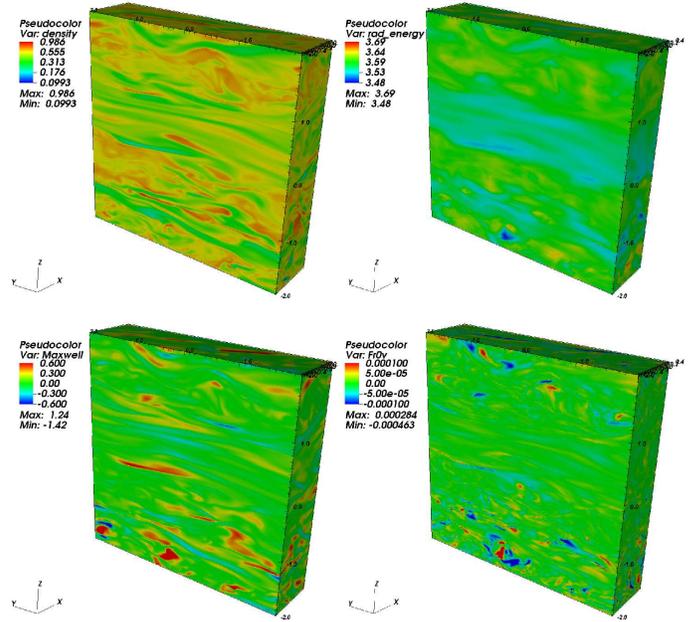}
\caption{Snapshots of  simulation YFR1Res at time $25.1$ orbits. From left to right, 
top to bottom, are density $\rho$, radiation energy density $E_r$, Maxwell stress 
$-B_xB_y$ and co-moving flux projected along the direction of perturbed velocity $(\bF_{r,0}\cdot\delta\bv)/|\delta\bv|$. }
\label{T1plots}
\end{figure}

\begin{figure}[htp]
\centering
\includegraphics[width=1.2\hsize]{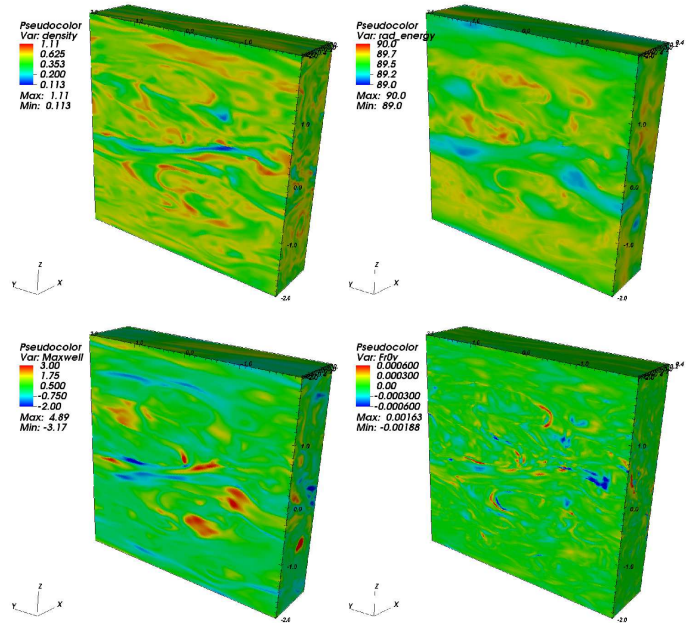}
\caption{The same as Figure \ref{T1plots} but for simulation YFR3Res at time $72.0$ orbits. }
\label{T3plots}
\end{figure}

\subsubsection{Time Evolution}
Temporal evolution of radiation and gas pressure, 
three components magnetic 
and kinetic energy densities as well as the total work 
done on the wall of the simulation box per unit time for 
simulation YFR3Res are shown in Figure \ref{YFT3ResHistory}. 
Because the initial radiation energy density is so large, within $100$ 
orbits, the total energy only increases $16.7\%$ due to the work done 
on the wall of the simulation box. Similar to case with net 
vertical flux, kinetic and magnetic energy densities saturate quickly 
after initial transient and fluctuate, although the total energy density never 
reaches an equilibrium value. 
%However, the fluctuations around the dynamical 
%time scales   

In Figure \ref{T1plots}, we show 
snapshots of $\rho$, $E_r$, $-B_xB_y$ and $(\bF_{r,0}\cdot \delta\bv)/|\delta\bv|$ 
for simulation YFR1Res at time $25.1$ orbits.  Here $\delta\bv$ 
is the perturbed velocity with the background shear subtracted. The same 
quantities for simulation YFR3Res at time $72.0$ orbits are shown 
in Figure \ref{T3plots}. Similar to the simulations in Table \ref{ZFtable}, 
YFR1Res and YFR3Res are also very compressible. The ratio between 
the maximum and minimum densities is $\sim 10$.  Most of the Maxwell 
stress is located in regions where the density variations are large, 
with the largest values in the low density regions. 
The radiation energy density has a similar distribution 
as density because $E_r$ is 
also compressed when $\rho$ is compressed. 
Therefore the co-moving flux $\bF_{r,0}$ is also large when 
the density gradient is large. Radiation energy 
density $E_r$ in 
simulation YFR3Res is $81$ times the value in simulations 
YFR1Res while the co-moving flux is almost a factor of $8$ larger 
in simulation YFR3Res than YFR1Res.  This means that the 
radiation force on the fluid in YFR3Res is also about $8$ times 
larger than the radiation force in YFR1Res. The sign of the quantity 
$(\bF_{r,0}\cdot\delta\bv)/|\delta\bv|$ represents the relative 
direction between the co-moving flux and the velocity field. When it is 
positive, photons accelerate the fluid during expansion while it is negative, 
photons decelerate the fluid 
during compression. The net effect is that part of the kinetic energy is 
converted to gas internal and radiation energy, resulting in damping 
due to radiation drag \citep{AgolKrolik1998}.  
The fact that co-moving flux in simulation YFR3Res is about $8$ times larger 
than the co-moving flux in simulation YFR1Res, implies that the radiation 
damping effect in YFR3Res will also be much larger than this effect in YFR1Res. 
The consequences of the radiation drag effect will be examined in more detail in
the following sections.

\begin{table}[htdp]
\centering
\begin{threeparttable}
\caption{Parameters for simulations with net toroidal flux}
\begin{tabular}{cccccccccc}
\hline
%Label & $\kappa_s$	& $\kappa_a$	&	$L_x$/$L_0$ & $L_y$/$L_0$ & $L_z$/$L_0$ & $P_{g}$ & $P_{r}$ & $B_{0}^2/2$ & $N$/$L_0$\\
Label & 	 $P_{g}$ & $P_{r}$ &  $N$/$L_0$\\
\hline
YFNR 	&   	6.37		&  ---		& 	32\\
YFR1 	&   	0.50			&  5.87	&	 32	\\	
YFR2 	&  	1.00			&  93.92	&  	 32	 \\
YFR3 	& 	1.50			& 475.47	&   	 32	\\
YFR1Res 	& 	0.50			& 5.87	&       64	\\
YFR2Res 	&   	1.00			& 93.92	&       64	\\
YFR3Res 	&     	1.50			& 475.47	&  	64	\\
%YFR1Thin	& $5.94\times10^3$  & $0.66$ 		& 1 &  4  	&  4	&  	0.5			& 5.88	&    $0.04$ 	& 32	\\
%VFR1	& $5.94\times10^3$  & $31.05$  	& 4 &  4  	&  1	&  	0.50			& 5.87	&    $3.1\times10^{-4} $ & 	  32	 \\
%VFR2	& $5.94\times10^3$  & $31.05$  	& 4 &  4  	&  1	&  	1.00			& 93.92	&    $3.1\times10^{-4} $ &   32 \\
%VFR3	& $5.94\times10^3$  & $31.05$  	& 4 &  4  	&  1	&  	1.50			& 475.47	&    $3.1\times10^{-4} $ &  32\\
%VFR2Res	& $5.94\times10^3$  & $31.05$  	& 4 &  4  	&  1	&  	1.00			& 93.92	&    $3.1\times10^{-4} $ &  64\\
%VFR3Res	& $5.94\times10^3$  & $31.05$  	& 4 &  4  	&  1	&  	1.50			& 475.47	&    $3.1\times10^{-4} $ &  64\\
%YFRZS	& $5.94\times10^3$  & $31.05$ 	& 1 &  4  	&  1	&       0.5			&  5.88	& $0.04 $ 			\\
%YFRLY 	& $5.94\times10^3$  & $31.05$  	&1 &  8  	&  1	&      0.5			&  5.88	 &  $0.04 $		\\
%YFNRLB 	& ---  & ---  	& 2 & 16 	&  2	& 	6.38			&  ---		 &  $2.0 $		\\
%YFRLB 	& $5.94\times10^3$  & $31.05$  	& 2 & 16 	&  2	& 	0.5 			& 5.88	 &  $2.0 $		\\
%YFRL  & $5.94\times10^3$   &  $31.05$  	& 2 & 16 	&  2	& 	0.5			& 5.88	 &  $0.04 $		\\
\hline
\end{tabular}
\label{YFtable}
\end{threeparttable}
\end{table}%

\subsubsection{Statistical Properties}
Averaged properties between $10$ and $100$ orbits for 
these simulations are shown in Table \ref{YFtable2}. By 
comparing YFR1, YFR2, YFR3 and YFR1Res, YFR2Res 
and YFR3Res, we see that when the initial temperature is $T_0$, 
the simulation with $32$ cells per $L_0$ is already 
converged, while for the simulations with initial 
temperatures $2T_0$ and $3T_0$, this is not the case. 
To check the convergence of the high temperature runs, 
we carry out another simulation with $128$ grids per 
$L_0$ for initial temperature $3T_0$. We only run this 
high resolution module for $40$ orbits. Averaging 
over the last $20$ orbits, $\<-B_xB_y\>=0.220$, 
$\<\rho v_x\delta v_y\>=0.00780$ and $\alpha_{\text{mag}}=0.258$. 
Comparing this run and simulations YFR3 and YFR3Res, we find that 
the differences between $64$ cells and $128$ cells per $L_0$ 
are much smaller than the differences between $32$ cells and 
$64$ cells per $L_0$. For example, the Maxwell and Reynolds stress 
only change within the statistical fluctuations when we  go from $64$  
to $128$ cells per $L_0$. Thus, we infer that the high temperature runs 
with $64$ cells per $L_0$ are approaching convergence. 
The analysis below will focus on the four simulations
YFNR, YFR1Res, YFR2Res and YFR3Res. 

\begin{table*}[htdp]
\caption{Temporal and Spatial Averaged Properties of Simulations with Net Azimuthal flux}
\centering
\begin{tabular}{cccccccccc}
\hline
Label & $\<-B_xB_y\>$ & $\<\rho v_x\delta v_y\>$ & $\<B^2/2\>$ & $\<B_x^2/2\>$	& $\alpha_{\text{stress}}$	 \\
\hline
YFNR		&	0.0330	&	0.00730		&	0.123	&	0.0122			&	4.52	\\				
YFR1		&	0.0469	&	0.00580		&	0.236	&	0.0169			&	8.09	\\			
YFR2		&	0.0737	&	0.00290		&	0.322	&	0.0237			&	25.4 \\	
YFR3		&	0.0547	&	$5.27\times 10^{-4}$		&	0.261	&	0.0129	&	104	\\	
YFR1Res		&	0.0425	&	0.00670		&	0.199	&	0.0175			&	6.34	\\
YFR2Res		&	0.125	&	0.00660		&	0.666	&	0.0462			&	18.9 \\			
YFR3Res		&	0.249	&	0.00570		&	1.01		&	0.0913			&	43.7 \\			
%YFRP1Thin	&	0.0273	&	0.00190		&	0.204	&	0.00540			&	\\
%VFR1		&	0.180	&	0.0197		&	0.418	&	0.0664			&	9.14 \\
%VFR2		&	0.251	&	0.0116		&	0.631	&	0.0858			&	21.6 \\
%VFR3		&	0.123	&	0.00160		&	0.427	&	0.032			&	76.9 \\
%VFR2Res		&	0.367	&	0.0166		&	0.904	&	0.128			&	22.1 \\
%VFR3Res		&	0.510	&	0.0103		&   	1.41 		&	0.179			&	49.5 \\
\hline
\hline
Label &  $\<B_y^2/2\>$ & $\<B_z^2/2\>$ & $\<T/T_0\>$  &  $\alpha_{\text{mag}} $	&	$\alpha $\\
\hline
YFNR		&	0.107	&	0.00410	&	3.03		&	0.267		&	0.00632	\\
YFR1		&	0.210	&	0.00920	&	1.21		&	0.199			&	0.00826	\\
YFR2		&	0.281	&	0.0175	&	2.01		&	0.160			&	0.000806	\\
YFR3		&	0.236	&	0.0117	&	3.02		&	0.210			&	0.000116	\\
YFR1Res		&	0.174	&	0.00790	&	1.20		&	0.214			&	0.00772	\\
YFR2Res		&	0.592	&	0.0279	&	2.09		&	0.188			&	0.00140 \\			
YFR3Res		&	0.861	&	0.0561	&	3.05		&	0.247			&	0.000534\\			
%YFRP1Thin	&	0.0273	&	0.00190		&	0.204	&	0.00540			&	\\
%VFR1		&	0.323	&	0.0285		&	1.45		&	0.431			& 0.0340	\\
%VFR2		&	0.500	&	0.0445		&	2.16		&	0.398			& 0.00279	\\
%VFR3		&	0.373	&	0.0221		&	3.02		&	0.288			& 0.000262	\\
%VFR2Res		&	0.717	&	0.0588		&	2.19		&	0.406			& 0.00407	\\
%VFR3Res		&	1.13	&	0.0996		&   	3.11		&	0.362			& 0.00109	\\
\hline
\hline
\end{tabular}
\label{YFtable2}
\end{table*}%

\subsubsection{Change of Maxwell stress with total pressure}
All the simulations done in this paper have fixed vertical 
box size and thus we cannot study the increase of Maxwell stress 
due to the increase of disk scale height, which we expect to be the primary 
source of connection between stress and pressure. This will 
be studied in future work. However, as Figure \ref{3THighres} shows, even 
with fixed box size, the Maxwell stress is increased with increasing 
total pressure. 
Compared with simulation YFR1Res with temperature $T=T_0$, 
initial total pressure is increased by 
a factor of $15$ when temperature $T=2T_0$ (run YFR2) and a factor of $75$ when temperature 
$T=3T_0$ (run YFR3). However, total stress is only increased by a factor of 2.67 from 
simulation YFR1Res to YFR2Res, and a factor of $5.18$ from simulation 
YFR1Res to YFR3Res. 

To identify the cause of increase in Maxwell stress as total pressure increases, we 
have studied the magnitudes of the radiation source terms in detail. 
In equation (\ref{equations}), the terms that are responsible for the radiation drag 
effect are the radiation momentum source term ${\bf S_r}(\bP)$ and the
velocity dependent radiation work terms in the  
radiation energy source term $S_r(E)$. The amplitude of these source terms 
for simulations YFR1Res, YFR2Res and YFR3Res can be measured directly 
from the simulation data. We first calculate the spatial 
and temporal averaged radiation momentum source term 
$S_p\equiv\<\Prat(\sigma_{aF}+\sigma_{sF})|\bF_{r,0}|\>$. The temporal 
average is taken over $10$ and $100$ orbits to avoid the effect of initial 
transients. For YFR1Res, YFR2Res and YFR3Res, in units of $\rho_0a_0\Omega$, $S_p$ are $1.01$, $3.90$ 
and $6.38$ respectively. 
As expected, the radiation momentum source term is increased 
when temperature and thus radiation pressure is increased. Second, we calculate $t_r=\<E_k\>/\<W_r\>$, the averaged 
ratio between the 
kinetic energy $E_k$ and the radiation work term $W_r\equiv -\Prat(\sigma_{aF}-\sigma_{sF})\bv\cdot\bF_{r,0}$. 
This is the time scale to change the kinetic energy due to the work done 
by the radiation force. Here $E_k$ does not include the kinetic energy due to background 
Keplerian motion. For simulations YFR1Res, YFR2Res and YFR3Res, in our time unit $\Omega^{-1}$, 
$t_r$ is $-5.70$, $-3.33$ and $-3.20$ respectively. 
%For simulation ZFRS, the number is -4.0 
Here $t_r$ is negative, which means that the 
co-moving flux prefers to point to the opposite direction 
of fluid motion and decelerates it on average.  
The time scale is shorter for simulations YFR2Res and YFR3Res, 
only about half an orbital period. To make sure that the momentum source term ${\bf S_r}(\bP)$ 
is the key to the increase the Maxwell stress, 
we have also carried out another test in which we remove the momentum source term ${\bf S_r}(\bP)$ 
and the radiation work terms in the radiation energy source 
terms. We use the same setup as simulation YFR3Res. We find almost the same Maxwell 
stress and Reynolds stress as simulation YFR1Res and YFNR. Therefore, if there is no damping 
of fluid motion by the photons, we find the same saturation level as in the pure gas pressure case. 
This is strong evidence that it is radiation damping that is responsible for the increase of the 
Maxwell stress.

The importance of the radiation drag effect can also be quantified by the photon 
bulk viscosity \citep[e.g.,][]{Castor2004}. 
In our units, the radiation bulk viscosity is (equation 6.80 of \citealt{Castor2004})
\begin{eqnarray}
\xi_r=4\Prat E_r/\left(9\sigma_t \Crat\right) ,
\label{Radviscosity}
\end{eqnarray} 
where $\sigma_t=\sigma_{aF}+\sigma_{sF}$. 
A Reynolds number can be defined based on the radiation bulk viscosity, 
and the typical velocity, length scale and density (isothermal sound speed $c_s$, scale 
height $c_s/\Omega$ and fiducial density $\rho_0$ respectively), that is 
\begin{eqnarray}
R_e=\rho_0 c^2_s /\left(\Omega\xi_r\right).
\end{eqnarray} 
For simulations YFR1Res, YFR2Res and YFR3Res, the Reynolds numbers are $1.91\times10^5$,  
$1.18\times10^4$ and $2.33\times10^3$ respectively. For simulation ZFRS, this 
number is $5.71\times10^4$. 

Our simulations also include grid scale dissipation set by numerical viscosity and resistivity. 
In order for the photons to control the 
magnetic dissipation rate, the Prandtl number defined by the ratio between radiation bulk 
viscosity and the numerical resistivity needs to be larger than $1$. Numerical resistivity is likely 
problem dependent. \cite{Simonetal2009} quantifies the effective magnetic Reynolds number 
based on numerical resistivity in Athena to 
be around $5\times10^3 - 1\times 10^4$. \cite{SimonHawley2009} also find that when 
the Reynolds number for the explicit (shear) viscosity is smaller than $6400$, the Maxwell stress 
begins to show a significant increase. Therefore we suggest that radiation pressure in simulations 
YFR2Res and YFR3Res is large enough to change the Maxwell stress significantly while 
in simulations YFR1Res and ZFRS, it is not.

A similar result for the unstratified disk model was
also reported for the pure gas pressure case by \cite{Sanoetal2004}. 
They studied the saturation level 
of MRI for a wide range of gas pressure and found that Maxwell stress 
has a weak dependence on the gas pressure. For the zero net flux case, they 
found $\<-B_xB_y\>\propto P^{1/4}$ while for the net vertical flux case, 
they found $\<-B_xB_y\>\propto P^{1/6}$. There is no definite 
explanation for the weak dependence of Maxwell stress on the gas pressure. 
They argue that higher gas pressure can reduce the magnetic 
reconnection rate and thus reduce the dissipation rate of magnetic energy, 
resulting in a higher Maxwell stress in the saturation state. 
More recently, studies with explicit viscosity and resistivity 
\citep[e.g.,][]{LesurLongaretti2007,FromangPapaloizou2007,SimonHawley2009,Simonetal2009}
found that at moderate 
Reynolds number $Re$, the Maxwell stress from the saturation states of 
MRI turbulence increases 
with the magnetic Prandtl number $Pm=\nu/\eta$, which 
is the ratio between microscopic viscosity and resistivity. 
Following \cite{BalbusHawley1998}, \cite{SimonHawley2009} argues 
that viscosity 
can prevent motion that would bring field together on small scales 
and therefore reduce the magnetic reconnection rate. 
The increase of Maxwell stress with radiation pressure found here, may be 
due to a similar mechanism, except that the change of effective Prandtl 
number is caused by the damping of radiation field on the compressible motions in the fluid, 
rather than explicit viscosity.

\subsubsection{The ratio between Maxwell stress and Reynolds stress}
\begin{figure}[htp]
\centering
\includegraphics[width=0.9\hsize]{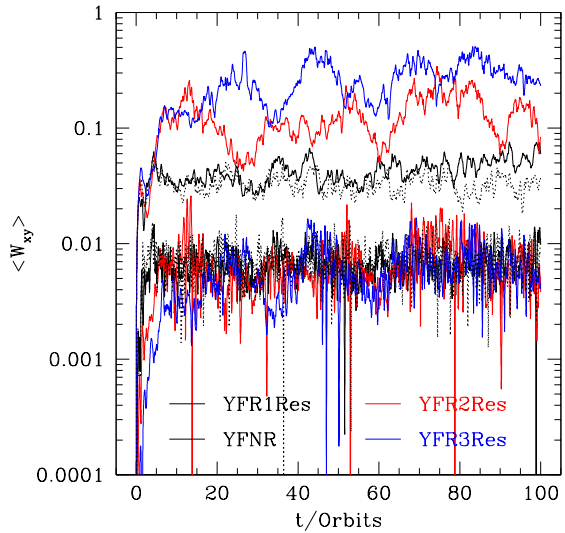}
\caption{ History of the spatially averaged Maxwell and Reynolds stress 
for simulations with net azimuthal flux. 
Maxwell stress is plotted as the upper thick lines while the Reynolds stress 
is plotted as the lower thin lines. 
Parameters of all the simulations shown in this figure are 
listed in Table \ref{YFtable}. YFR1Res, YFR2Res and YFR3Res differ in the initial ratio between radiation 
pressure and gas pressure, which increases by a factor of $27$ from YFR1Res to YFR3Res. YFNR is 
a comparison simulation without radiation field but the same total pressure as YFR1Res.}
\label{3THighres}
\end{figure}

The change in the effective Prandtl number due to radiation drag also 
seems to affect the ratio of the Maxwell to Reynolds stress. 
In Figure \ref{3THighres}, when the Maxwell stress is increased with 
increasing radiation pressure, the Reynolds stress stays at almost the 
same level. Therefore, the ratio between Maxwell 
stress and Reynolds stress is increased when radiation pressure is 
increased. As calculated in Table \ref{YFtable}, 
this ratio is $6.34$ for YFR1Res and it increases to $18.9$ and $43.7$
for YFR2 and YFR3 respectively. As a comparison, the ratio 
is only $4.52$ for the simulation YFNR without radiation. This ratio 
reported in most MRI simulations with or 
without radiation is also around $4-5$ \citep[e.g.,][]{HGB1995,Turneretal2003,Sanoetal2004,Davisetal2010}. 
However, when \cite{SimonHawley2009} includes explicit viscosity 
in their MRI simulations, they finds that the ratio between Maxwell stress 
and Reynolds stress increases when the stress and viscosity is increased, 
although the change of this ratio is not as large as what we find here. 

In the test run in which we remove the radiation momentum source term 
described in the last section, we find the usual ratio between Maxwell 
stress  and Reynolds stress as in the pure gas pressure case. We also notice 
that for all the simulations listed in Table \ref{ZFtable}, where we 
find the radiation damping is not important, the ratio between Maxwell stress 
and Reynolds stress is also normal. This ratio is only significantly increased 
for those simulations listed in Table \ref{YFtable}, where the Maxwell stress is 
much larger than what the values would be if the radiation pressure is replaced 
with the same amount of gas pressure. Those results suggest that the 
change of the ratio between Maxwell stress and Reynolds stress is due to the same 
physical mechanism responsible for the increase of Maxwell 
stress discussed in the last section, which is the damping of 
the compressive motions due to radiation drag. 
We have also checked that this result is independent of 
field geometry. When we use net vertical magnetic flux, 
we get the same result.

We also notice that $\alpha_{\text{mag}}$ for those simulations listed in 
Table \ref{YFtable} is systematically smaller than $\alpha_{\text{mag}}$ 
from non-radiative MRI simulations. It is also smaller than 
$\alpha_{\text{mag}}$ for the simulations listed in Table \ref{ZFtable}, 
which have net vertical magnetic flux. 
 For simulations YFR1Res, YFR2Res and YFR3Res, $\alpha_{\text{mag}}$
varies from $0.188$ to $0.247$ with the average value to be $0.21$. 
For simulations ZFRS, ZFRL, ZFNRS, ZFNRL and ZFRS64,  
$\alpha_{\text{mag}}$ is always between 0.4 and 0.5, which is also the 
usual value reported by most non-radiative MRI simulations \citep[e.g.,][]{Hawleyetal2011}. 
A smaller $\alpha_{\text{mag}}$ is also found from stratified simulations of radiation dominated 
accretion disk done with ZEUS (Krolik et al., 2012, private communication).

\subsubsection{Radiative Viscosity}
In the radiation dominated flow around the black holes, it has been argued 
that radiative viscosity can be important to transfer the angular momentum, 
which can even be the dominant mechanism for spherical accretion flows \citep[e.g.,][]{LoebLaor1992}. 
Radiative viscosity will only be important when the radiation energy 
density is significant. Here, we can compare the viscous stress due to radiation with 
the Maxwell and Reynolds stress for simulation YFR3Res, which has the largest 
radiative viscosity in our simulations. 

In principle, the off-diagonal component of the co-moving radiation pressure 
in the $x-y$ plane, which is responsible for the transfer of angular momentum, 
comes from two parts in our formula. 
The first part is the off-diagonal component of the Eddington tensor in the Eulerian frame. 
We have calculated the VET for simulation YFR3Res with the short-characteristic module, 
and find that the off-diagonal components of the VET is smaller than $10^{-8}$. Thus 
it can be neglected. The second 
part is the radiative shear viscosity, the formula of which is \citep[e.g.,][]{Castor2004}
\begin{eqnarray}
P_{r,xy}=-\frac{4\Prat E_r}{15\sigma_t \Crat}\left(\frac{\partial v_x}{\partial y} + \frac{\partial v_y}{\partial x}\right).
\end{eqnarray}
This is caused by the velocity dependent radiation momentum source terms. 
For simulation YFR3Res, the spatially and timely averaged value is $\<P_{r,xy}\>=4.13\times10^{-4}$ 
in our fiducial units. The radiative viscosity is dominated by the background shearing. 
Therefore,  even for simulation YFR3Res, the off-diagonal component of the co-moving 
radiation pressure is only $1.66\times10^{-3}$ of the Maxwell stress and $7.2\%$ of 
the Reynolds stress. 

%\begin{figure}[htp]
%\centering
%\includegraphics[width=0.48\hsize]{YF3T_lowRes.ps}
%\includegraphics[width=0.48\hsize]{Vert3T_lowRes.ps} 
%\caption{\emph{Left:} History of the spatial averaged Maxwell stress and Renold stress 
%for the simulations with net azimuthal flux. 
%Maxwell stress is plotted as the upper level thin lines while the Reynolds stress 
%is plotted as the lower level thick lines. 
%Parameters of all the simulations shown in this figure are 
%listed in Table \ref{YFtable}. YFR1, YFR2 and YFR3 different from the initial ratio between radiation 
%pressure and gas pressure, which increases by a factor of $27$ from YFR1 to YFR3. YFNR is 
%a comparison simulation without radiation field but the same total pressure as YFR1. 
%\emph{Right:} The same as the left panel but for the simulations with net vertical flux. 
% }
%\label{3Tlowres}
%\end{figure}

\section{Discussion and Conclusion}
\label{sec:discussion}
With our recently developed Godunov radiation MHD 
code, we have confirmed the results of 
\cite{Turneretal2003} for unstratified optically 
thick mid-plane of a radiation dominated accretion disk. 
In particular, we find when the photon diffusion length per 
orbit is larger than the most unstable MRI wavelength, 
the MRI turbulence becomes very compressible. 
Furthermore, we extend the fiducial simulations of
\cite{Turneretal2003} by increasing the radial size by a factor of $4$, 
which allows more radial modes to exist in the simulation box. 
The saturation level of the MRI 
from the larger simulation domains is very similar to what we find in the 
fiducial runs. Recurrent channel solutions observed in small 
box simulations do not exist when radial size is increased. This is consistent 
with what \cite{Bodoetal2008} found without radiation. In this sense, 
radiation pressure plays the same role as gas pressure in the optically 
thick regime. 

When the radiation energy density is increased to a non-negligible 
fraction ($\gtrsim 1\%$) of fluid rest mass energy, we find that Maxwell 
stress is increased with increasing radiation pressure. 
At the same time, Reynolds stress seems to be independent of 
radiation pressure. Therefore, the ratio between 
Maxwell and Reynolds stress is increased with increasing 
radiation pressure. We suggest that this is due to the damping effect 
of radiation drag which introduces an effective bulk viscosity $\xi_r$. 
To quantify the importance of the radiation damping 
effect, we have calculated the Reynolds number based on $\xi_r$ (equation \ref{Radviscosity}) 
and find values of $\sim 2.33\times10^3$ for the simulation 
with the largest radiation pressure. 
This number depends on $\Prat E_r/\Crat$ in our units for a fixed opacity, which 
is actually the momentum of the photons. This can explain why we do not 
see the change of Maxwell stress as well as the ratio between Maxwell 
and Reynolds stress compared 
to the pure gas pressure case for the simulations listed in Table \ref{ZFtable}. 
%The opacity $\sigma_t$ 
%is very similar for the simulations in Table \ref{ZFtable} and \ref{YFtable}. 
Although the 
ratio between radiation pressure and gas pressure is $125$ for the simulations 
listed in Table \ref{ZFtable}, $\Prat E_r/\Crat$ is only $7.7\%$. While for the simulations 
listed in Table \ref{YFtable}, $\Prat E_r/\Crat$ is $0.035$, $0.567$ and $2.87$ 
for $T=T_0, 2T_0, 3T_0$ respectively.

Because there is no cooling in the unstratified simulations, the disk is continuously 
heated up due to the dissipation of MRI. However, as the histories of these 
simulations show (Figure \ref{Turnerstress} and Figure \ref{3THighres}), 
the secular build-up of energy in these simulations does not affect the diagnostics 
we are interested, such as Maxwell 
stress and Reynolds stress. This is not surprise, because the effects discussed this 
in paper requires $T$ to increase by a factor of $2$ and $3$, while during the $100$ 
orbits of YFR1, YFR2Res and YFR3Res, $T$ is only increased by $40\%$, $9\%$ 
and $4\%$ respectively.

Even when radiation energy density reaches $1\%$ of the fluid rest mass energy density as 
in simulation YFR3Res, we find that the off-diagonal component of the co-moving radiation pressure 
is much smaller than the Maxwell and Reynolds stress. This confirms that 
transfer of angular momentum by radiative viscosity is not the dominant 
mechanism when the turbulence due to MRI exists. 

%Second, dynamo process is believed to be responsible for the 
%butterfly pattern observed in all the stratified MRI simulations 
%\citep[e.g.,][]{Brandenburgetal1995,Gressel2010,Blackman2012}. 
%As dynamo process requires the interactions between velocity field 
%and magnetic field in a way such that $B_x$ and $B_y$ can regulate 
%each other, the change of the typical scale of the velocity field 
%by the photons may eventually change the dynamo process and thus 
%the butterfly patter. As the butterfly patter is closely related to 
%the magnetic buoyancy, which takes the magnetic field away from 
%the mid-plane to the photosphere, we thus expect that the distribution 
%of magnetic field and the vertical structure of the accretion disk 
%will eventually be affected by the strong radiation field. 
%But we only expect this to happen when the radiation energy density 
%in the mid-plane is a significant fraction of the rest mass energy of the fluid. 
%This will be studied in the future 
%with simulations of stratified disk models. 

We perform two sets of simulation with both high ($N/L_0=64$) and low
($N/L_0=32$) resolution.  We find consistent (statistically equivalent)
estimates for turbulent properties, such as the stress, which suggests these
simulations are converged.  Since we do not include explicit resistivity,
our results may have been sensitive to the effective amount of ``numerical
resistivity", which is strongly resolution dependent \citep[e.g.,][]{Simonetal2009}.
Our result implies that effective resistivity is sufficiently small that
turbulent properties are set primarily by dynamics at larger scales and
not strongly impacted by the precise properties of grid scale
dissipation.

In all the simulations, we have fixed the the Eddington tensor to be $1/3
\delta_{ij}$ in the Eulerian frame.  Since our simulations are
approximating the mid-plane of an optically thick accretion disk, the
assumption of isotropy is well justified, but it is generally expected
that the radiation field will be nearly isotropic in the {\it co-moving}
frame.  As the difference between co-moving radiation pressure 
and Eulerian radiation pressure in our units is $\bv\cdot \bF_{r,0}/\Crat$ \citep[e.g.,][]{MihalasMihalas1984},  
the difference between Eulerian Eddington tensor and co-moving Eddington 
tensor will be $\bv\cdot\bF_{r,0}/(\Crat E_r)$, which we confirm is less than $10^{-4}$ for all the simulations presented
here.  Therefore, this assumption should not appreciably impact our
results. A test with Eddington tensor to be $1/3\delta_{ij}$ in the co-moving 
frame does confirm this conclusion. 

We have also calculated VET with the short characteristic module described in 
\cite{Davisetal2012} for the simulations shown in Table \ref{ZFtable}. For the 
unstratified shearing box disk models, there is no preferred direction 
for the specific intensity because periodic or shearing periodic boundary conditions 
are applied for all three directions. Furthermore, photon mean free paths for all 
the simulations are smaller than the cell size. Therefore, VET returned by 
the short characteristic module only differs from $1/3\bI$ by $10^{-4}$, 
when $24$ angles are used. This confirms that Eddington approximation 
is valid for the unstratified disk models, which are designed to study the optically 
thick mid-plane of accretion disks. 

The increase of Maxwell stress with increasing radiation pressure 
and the change of $\alpha_{\text{stress}}$ have important
implications. This study confirms that the damping effect 
of large radiation pressure will not decrease the total 
stress. Instead, the total stress is actually increased slowly with total pressure 
when radiation energy density is significant.

These results also confirm the important role radiation viscosity
plays as a source of microscopic dissipation in radiation dominated
regions of accretion flows.  Although radiation viscosity always plays
a subdominant role to Maxwell and Reynolds stresses in angular
momentum transport, it will generally exceed the standard plasma
viscosity.  In cgs units, Equation (\ref{Radviscosity}) corresponds to a
kinematic viscosity with
\begin{eqnarray}
\nu_r \simeq \frac{4a_r T^4}{9\kappa_{\rm T} \rho^2 c} = 3.4 \times 10^{-25} \frac{T^4}{\rho^2}\ {\rm cm}^2\ {\rm s}^{-1}.
\end{eqnarray} 
This can be compared with Equation (2) of \cite{BalbusHenri2008}, which gives an
estimate of the plasma viscosity $\nu_p\propto T^{5/2}/\rho$.  In the radiation dominated
regimes of accretion flows, $\nu_r > \nu_p$ when $T\lesssim 10^9$ K
and it is often the case that $\nu_r \gg \nu_p$. It is still true for the hottest radiation 
dominated flows ($T\gtrsim 10^9$ K) when $\rho<4.74\times10^4\left(T/10^{9}\ \text{K}\right)^{3/2}$ g cm$^{-3}$. 
% in all but the hottest
%flows ($T \gtrsim 10^8 K$) 
So radiation is the dominant viscosity associated with
microscopic dissipation. 
This makes radiation dominated accretion disk one
of the few regimes where the physically relevant value of the viscosity can
be numerically resolved with current computing resources. 
The ratio $\nu_r/\nu_p$ is larger 
for the accretion disks around supermassive black holes than 
solar mass black holes, which may lead to different properties of 
the accretion flows for systems in the two different scales if the 
radiation viscosity becomes important.

Despite the dominance of radiative viscosity, it is generally true
that plasma resistivity dominates over radiative resistivity, even
in radiation dominated accretion flows \citep{AgolKrolik1998}.  Therefore,
Equation (1) of \cite{BalbusHenri2008} still gives the relevant magnetic resistivity
$\eta$.  Since this quantity is generally very small compared with
$\nu_r$, the corresponding magnetic Prandtl number ${\rm Pm} =
\nu_r/\eta \gg 1$ in radiation dominated flows.  This conclusion is
interesting in light of speculation that Pm may be the primary parameter
that controls angular momentum transport by the MRI
\citep{Fromangetal2007}. A large Pm would place radiation dominated
accretion flows firmly in the efficient angular momentum regime if
this supposition is correct.

In all the unstratified disk simulations, the total energy can never 
reach an equilibrium value because there is no cooling in these 
simulations. However, the Maxwell and Reynolds stress saturate 
quickly after an initial transient. Even when total pressure is increased 
by a factor of $2$, the stress does not show any systematic trend 
of change within the first $100$ orbits of the simulations. 
This is because the saturation level in the unstratified 
disk models is limited by the box size. As found by \cite{HGB1995}, 
Maxwell stress in the saturated state increases linearly with 
box size in their unstratified disk models. Therefore, 
the unstratified disk models cannot be used to check the assumed relation 
between stress and pressure in the $\alpha$ disk model, which 
requires the change of disk scale height to connect them. 
The fact that the effective $\alpha$ value decreases when total pressure 
increases for these unstratified simulations is an artifact of the 
closed-box assumption, which should not be interpreted literally. 
When the total pressure increases,  
the disk scale height will also be increased, which gives more 
space for the turbulence to develop. 
The weak dependence of Maxwell stress on the radiation pressure 
due to the mechanism suggested here 
is probably not the dominant 
process in real accretion disks. More realistic stratified disk models will be studied in 
the future. Nevertheless, the phenomenon 
we find here is also likely to operate in the mid-plane of the accretion 
disk near the Eddington limit.
% It is also interesting to see 
%how the properties of the accretion flow will be affected by this effect. 
In the future, we will perform more realistic
stratified disk models that will be better suited for addressing these
questions.

\section*{Acknowledgements}
We thank Julian Krolik, Omer Blaes, Shigenobu Hirose and Jeremy Goodman for helpful 
discussions on the results. We also thank the referee, Chris Reynolds, for valuable 
comments to improve this paper.
 This work was supported by the
NASA ATP program through grant NNX11AF49G, and by computational resources
provided by the Princeton Institute for Computational Science and Engineering. 
Some of the simulations are performed in the Pleiades Supercomputer provided by 
NASA. 
This work was also supported in part by the U.S. National Science
Foundation, grant NSF-OCI-108849. 
% This last line is required as we use John Hopkins's cluster 
SWD is grateful for financial support from 
the Beatrice D. Tremaine Fellowship.
 
%\clearpage
 
\begin{appendix}
\section{Improvements to the Numerical Algorithm} 
\label{codechange}
The code used in this paper is based on the 
algorithm described in JSD12, with 
several important improvements. Based on our tests, those 
improvements are necessary to reduce numerical diffusion 
in the optically thick regime while leave the optically thin 
regime unaffected. 
%Those improvements do not change 
%the results shown in JSD12. Thus, here we 
%only show some test problems for the regimes that are not 
%covered by JSD12.

\subsection{Effective photon propagation speed}
The first change is the HLLE Riemann solver for the 
radiation subsystem (the last two lines of equation \ref{equations}). 
As described in \S 3.3.2 of JSD12, the HLLE Riemann 
solver is used to calculate the flux for the radiation subsystem. 
In JSD12, the characteristic speed along each direction is 
always chosen to be $\sqrt{f} \Crat$, where 
$f$ is the component of the Eddington tensor along that direction, 
following \cite{SekoraStone2010} (see their equation 46).  
However, as discussed by \cite{Auditetal2002}, when the optical depth per 
cell is larger than one, the numerical diffusive flux due to the HLLE Riemann solver 
can be much larger than the physical radiative flux, which can cause 
unphysical solutions. This issue does not appear in the tests done by 
\cite{SekoraStone2010} because they only consider pure absorption opacity. 
In the optically thick regime, the evolution of radiation energy density is dominated 
by the source term $\Crat\sigma_a(T^4-E_r)$ because the thermalization time is usually 
very short. However, for scattering opacity dominated cases, the energy source 
term is very small and the radiation flux term $\bfnabla\cdot \bF_r$ controls the change 
of $E_r$. If the numerical diffusive flux is much larger than the physical radiation flux, the photon
diffusion time will be significantly underestimated.

\cite{Auditetal2002} fixes this issue by changing the term $\bfnabla\cdot {\sf P_r}$ in their dimensionless 
units (see their equation 46). This is equivalent to reducing the characteristic speed of the radiation modes in the optical 
thick regime. To find the appropriate effective speed for different frequency mean opacity 
$\sigma_t\equiv\sigma_{aF}+\sigma_{sF}$, we carry out a linear analysis for the following 1D equations 
\begin{eqnarray}
\frac{\partial E_r}{\partial t}+\mathbb{C}\frac{\partial F_r}{\partial x}&=&0, \nonumber \\
\frac{\partial F_r}{\partial t}+f\mathbb{C}\frac{\partial E_r}{\partial x}&=&-\mathbb{C}\sigma_tF_r.
\label{effectivespeed}
\end{eqnarray}
The background state is a static medium with a uniform radiation field. For mode with wavenumber 
$k$, the propagation speed is 
\begin{eqnarray}
C_{eff}=\sqrt{f}\Crat\sqrt{1-\frac{\sigma_t^2}{4f k^2}}.
\label{reduceC}
\end{eqnarray}
In the optically thin limit when $\sigma_t\rightarrow 0$, the effective 
speed of light $C_{eff}$ approaches $\sqrt{f}\Crat$. In the optically thick 
limit, $C_{eff}$ is much smaller than $\sqrt{f}\Crat$. Especially, for modes 
with $\sigma_t^2/(4fk^2)>1$, the propagation speed of the photons
is zero, which means photons only diffuse through the medium instead of free-streaming. 
Because we solve the radiation subsystem implicitly, the distance that
photons can travel for a give time step $dt$ is not limited by the size of each cell. 
The lower limit for the wave number $k$ should be $1/(\Delta t \Crat)$. However, this 
value is usually too restrictive and makes the resulting matrix from the implicit 
radiation subsystem very hard to invert. 
Instead, another characteristic length is the size of 
each cell. Based on our tests, we find that if we calculate the effective propagation 
speed with a wave length of ten cells, the numerical diffusion is small enough not 
to affect the solution.  If the size of each cell is $\Delta l$, 
to connect the optically thin and optically thick regimes smoothly, 
the effective propagation speed we use is
\begin{eqnarray}
C_{eff}^{\star}&=&\Crat\sqrt{\frac{f}{\tau}\left[1-\exp{\left(-\tau\right)}\right]}, \nonumber \\
\tau&\equiv&(10\Delta l\times\sigma_t)^2/\left(2f\right).
\label{actualc}
\end{eqnarray}

In the optically thin limit, the effective speed is reduced to equation \ref{reduceC} 
with wavenumber $k=1/(10\Delta l)$. In optically thick limit, the effective speed 
$C_{eff}\propto 1/(\Delta l\times\sigma_t)$. The purpose of using $C_{eff}^{\star}$ instead of $\Crat$ 
is to reduce the numerical diffusion so that we can capture the physical radiation 
flux, which is small in the diffusion regime. As long as physical radiation flux is dominant 
over the numerical flux in all regimes, the solution is more accurate. 
Tests presented below do confirm this. 

\subsection{Advection Flux}
\label{AdvEr}
In the optically thick regime, the diffusive flux is very small, and the change of radiation 
energy density can be dominated by the advection flux. We find that the first order 
treatment for the radiation subsystem described in JSD12 is too diffusive for the advection 
flux. In order to improve the accuracy, we separate the advection flux from the diffusive flux 
and treat the advection flux explicitly with second order accuracy. 

For a given velocity $\bv$, the change of $E_r$ due to the advection flux is 
\begin{eqnarray}
\frac{\partial E_r}{\partial t}+\bfnabla\cdot(\bv E_r)=0.
\end{eqnarray}
 Because the radiation subsystem is separated from the MHD part, the flow velocity $\bv$ 
 is taken to be constant during this step. We first calculate the flow velocity at the 
 edge of each cell $V_{i+1/2}$ by taking the average of cell centered velocities in the two 
 neighboring cells.
We use the van Leer slope (equation 48 and 49 of \citealt{Stoneetal1992I}) 
to interpolate $E_r$ from cell centers to cell edges $E_{r,i+1/2}$. Then the advection flux on each cell edge 
$\mathcal{F}_{i+1/2}$ is just $\mathcal{F}_{i+1/2}=V_{i+1/2}E_{r,i+1/2}$.

\subsection{The New Implicit Backward-Euler Step}
In order to incorporate the above two improvements  into our implicit Backward Euler step, 
we first rewrite the radiation subsystem into the following form
\begin{eqnarray}
 \frac{\partial E_r}{\partial t}+\mathbb{C}\bfnabla\cdot \left(\bF_r-\frac{\bv E_r+\bv\cdot{\sf  P} _r}{\Crat}\right)+\bfnabla\cdot\left(
 \bv\cdot{\sf P}_r\right)&=&-\bfnabla\cdot\left(\bv E_r\right)+\mathbb{C}S_r(E), \nonumber \\
\frac{\partial \bF_r}{\partial t}+\mathbb{C}\bfnabla\cdot{\sf P}_r&=&\mathbb{C}{\bf S_r}(\bP).
\end{eqnarray}
Notice that $\bF_{r,0}=\bF_r-\left(\bv E_r+\bv\cdot {\sf P}_r\right)/\Crat$ is actually the co-moving flux. In this way, 
the advection part $\bfnabla\cdot\left(\bv\cdot {\sf P}_r+\bv E_r\right)$ is separated from the flux in the 
co-moving frame $\bfnabla\cdot \bF_{r,0}$. The advection part $\bfnabla\cdot\left(\bv\cdot {\sf P}_r+\bv E_r\right)$ 
is the advective radiation enthalpy \citep[e.g.,][]{Castor2004}. Only the $\bfnabla\cdot\left(\bv E_r\right)$ part 
is treated as described in \S\ref{AdvEr}, while the $\bfnabla\cdot\left(\bv\cdot {\sf P}_r\right)$ part is solved implicitly with the 
source terms.

First, with the velocity field $\bv$ and $E_r^n$ at time step $n$, we calculate the advection flux along each direction 
$\mathcal{F}_{i+1/2,j,k}$, $\mathcal{F}_{i,j+1/2,k}$, $\mathcal{F}_{i,j,k+1/2}$ as described in \S\ref{AdvEr} for each cell. 
Then we calculate the implicit HLLE fluxes ${\sf \bold{F}^{\text{HLLE}}}$ according to equation (39) of \cite{SekoraStone2010} 
for the following equation
\begin{eqnarray}
 \frac{\partial E_r}{\partial t}+\mathbb{C}\bfnabla\cdot \left(\bF_r-\frac{\bv E_r+\bv\cdot{\sf  P} _r}{\Crat}\right)&=&0, \nonumber \\
\frac{\partial \bF_r}{\partial t}+\mathbb{C}\bfnabla\cdot{\sf P}_r&=&0.
\end{eqnarray}
The following changes are made during the calculation of the HLLE fluxes.
The maximum and minimum wave speeds we  actually use are $C_{eff}^{\star}$ and $-C_{eff}^{\star}$ respectively, 
 where $C_{eff}^{\star}$ is calculated according to equation \ref{actualc}. In order to make the implicit matrix easier to invert 
 for the cases with sharp opacity jump, the opacity $\sigma_t$ we use to calculate 
$C_{eff}^{\star}$ on the cell edges is the average opacity of two neighboring cells. The term $\bfnabla\cdot\left(\bv\cdot{\sf P}_r\right)$ 
is calculated implicitly as
\begin{eqnarray}
\bfnabla\cdot\left(\bv\cdot{\sf P}_r\right)(i,j,k)&=&\left[\bv\cdot{\sf P_r^{n+1}}(i+1)-\bv\cdot{\sf P_r^{n+1}}(i-1)\right]/(2\Delta x)\nonumber \\
&+&\left[\bv\cdot{\sf P_r^{n+1}}(j+1)-\bv\cdot{\sf P_r^{n+1}}(j-1)\right]/(2\Delta y) \nonumber \\
&+&\left[\bv\cdot{\sf P_r^{n+1}}(k+1)-\bv\cdot{\sf P_r^{n+1}}(k-1)\right]/(2\Delta z).
\end{eqnarray}
Then radiation energy density $E_r$ is updated as 
\begin{eqnarray}
E_{r,i,j,k}^{n+1}&=&E_{r,i,j,k}^n-\frac{\Delta t}{\Delta x}\left[{\bf F}^{HLLE}_{i+1/2,j,k}-{\bf F}^{HLLE}_{i-1/2,j,k}\right]
-\frac{\Delta t}{\Delta y}\left[{\bf F}^{HLLE}_{i,j+1/2,k}-{\bf F}^{HLLE}_{i,j-1/2,k}\right]\nonumber\\
&-&\frac{\Delta t}{\Delta z}\left[{\bf F}^{HLLE}_{i,j,k+1/2}-{\bf F}^{HLLE}_{i,j,k-1/2}\right]
-\frac{\Delta t}{\Delta x}\left[\mathcal{F}_{i+1/2,j,k}-\mathcal{F}_{i-1/2,j,k}\right] \nonumber \\
&-&\frac{\Delta t}{\Delta y}\left[\mathcal{F}_{i,j+1/2,k}-\mathcal{F}_{i,j-1/2,k}\right]
-\frac{\Delta t}{\Delta z}\left[\mathcal{F}_{i,j,k+1/2}-\mathcal{F}_{i,j,k-1/2}\right] \nonumber \\
&-&\Delta t\bfnabla\cdot\left(\bv\cdot{\sf P}_r\right)
+\tilde{S}_r(E).
\end{eqnarray}
The special treatment of the energy source term $\tilde{S}_r(E)$ described in JSD12 is also used 
here in order to reduce the energy error. The implicit equation for radiation flux $\bF_r$ is 
unchanged.

\subsection{Numerical Tests of the Improved Algorithm}

We find that the most useful tests to demonstrate the necessity of those improvements 
is the dynamic diffusion test with pure \emph{ scattering} opacity. The improvements 
do not change the results of other tests done by \cite{SekoraStone2010} and JSD12, which 
will not be repeated here. With pure scattering opacity, the radiation energy source 
term is zero and the change of radiation energy density is totally controlled by the divergence 
of radiation flux term. Inaccurate treatment of this term will easily show up in this case. In contrast to 
the static diffusion case, the dynamic diffusion test also requires accurate treatment of the advection term. 
%Notice that \cite{SekoraStone2010} also did the static diffusion tests with pure \emph{absorption} 
%opacity and a small enough time step to resolve the light crossing time. 
%In that case,  thermalization time is very short in optically thick regime and the solution is 
%dominated by the source term. That is why the trouble caused by the inaccurate characteristic 
%speed in optically thick regime doesn't show up. 

\begin{figure}[htp]
\centering
\includegraphics[width=1.1\hsize]{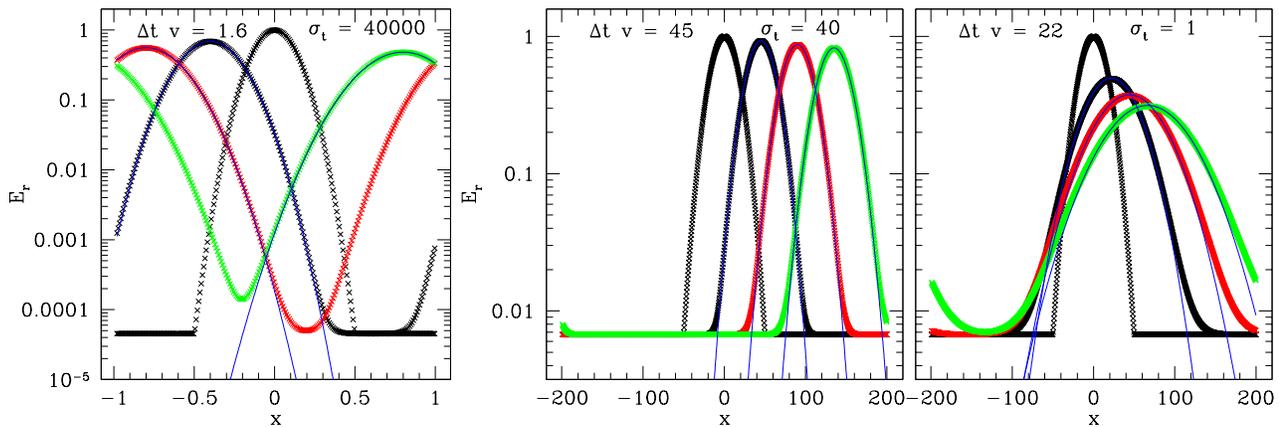}
\caption{Dynamic diffusion test for the radiation subsystem with our improved 
algorithm. From left to right, the parameters 
for equation \ref{diffusionsolution} are: $\sigma_t=40000,\nu^2=40$; $\sigma_t=40,\nu^2=0.002$; 
$\sigma_t=1,\nu^2=0.002$. The initial gaussian profiles are all centered at $x=0$. 
The black dots not centered at $x=0$, the red dots and the 
green dots are solutions from earlier time to later times with time intervals shown in each panel. 
The blue lines are analytical solutions (equation \ref{diffusionsolution}) at corresponding 
times, which agree with the analytical solution very well for the regions not close to the boundaries.
}
\label{diffusiontest}
\end{figure}

In the optically thick regime, the radiation subsystem can be approximated by diffusion equations. 
In 1D with a constant opacity $\sigma_t$ and the Eddington approximation, the diffusion equations 
and the analytic solution with initial Gaussian pulse for a static medium are given by equations (101) of \cite{SekoraStone2010}, 
except that $\sigma_t$ only contains scattering opacity here. For the dynamic diffusion test, we let the 
background fluid move with a constant velocity $v$. In the co-moving frame, the diffusion solution 
is the same as in the static diffusion case. In the Eulerian frame, the solution is
\begin{eqnarray}
E_r(x,t)=\frac{1}{\left(4Dt\nu^2+1\right)^{1/2}}\exp\left(\frac{-\left(\nu(x-vt)\right)^2}{4Dt\nu^2+1}\right),
\label{diffusionsolution}
\end{eqnarray}
with the initial condition 
\begin{eqnarray}
E_r(x,0)=\exp\left(-\nu^2 x^2\right),\quad\quad F_{r}(x,0)=\frac{2\nu^2x}{3\sigma_t}E_r(x,0)+\frac{4v}{3\Crat}E_r(x,0).
\end{eqnarray}
Here $D\equiv\Crat/(3\sigma_t)$ is the diffusion coefficient and $\nu$ is a free parameter to control the 
shape of initial Gaussian pulse. 
We can initialize the Gaussian pulse in the code and compare the analytical and numerical solutions 
at time $t$. We use periodic boundary conditions so that we do not need very large simulation box when the pulse 
is advected. The following parameters are chosen for 
all the tests: $\Crat=514.4$, $v=1$. We test the new algorithm in three cases with 
different opacities $\sigma_t=40000, 40, 1$. For the first case, 
the pulse is located within $|x|<0.5$ initially while for the other two cases, the pulse 
is located within $|x|<50$ initially. Results are shown in 
Figure \ref{diffusiontest}, which confirms that the 
improved algorithm can reproduce the diffusion behavior accurately. We have also 
tried that if we do not use those improvements, the diffusion time is too short compared 
with what it should be for the pure scattering case.

\end{appendix}

%\begin{thebibliography}{99}

%\end{thebibliography}
 
\bibliographystyle{apj}
\bibliography{NonStratDisk}
 
 \end{CJK*}
\end{document}